\documentstyle[epsf]{l-aa.pr}

\def\hi{\relax \ifmmode {\rm H\,{\sc i}}\else H\,{\sc i}\fi}
\def\hii{\relax \ifmmode {\rm H\,{\sc ii}}\else H\,{\sc ii}\fi}
\def\rq{$r^{1/4} $ }
\def\rh{$r^{1/2} $ }
\def\muo{\relax \ifmmode \mu_0\else $\mu_0$\fi}
\def\mue{\relax \ifmmode \mu_{\rm e}\else $\mu_{\rm e}$\fi}
\def\re{\relax \ifmmode r_{\rm e}\else $r_{\rm e}$\fi}
\def\magarc{mag arcsec$^{-2}$}
\def\boundboxo{\epsfbox[30 190 538 540]}
\def\boundboxt{\epsfbox[10 190 518 540]}

\def\abst{
 The statistics of the fundamental bulge and disk parameters of
galaxies and their relation to the Hubble sequence were investigated by
an analysis of optical and near-infrared observations of 86 face-on
spiral galaxies. The availability of near-infrared $K$ passband data
made it possible for the first time to trace fundamental parameters
related to the luminous mass while hardly being hampered by the effects
of dust and stellar populations.  The observed number frequency of
galaxies was corrected for selection effects to calculate volume number
densities of galaxies with respect to their fundamental parameters. The
main conclusions of this investigation are:\\ 1)\ \  Freeman's law has
to be redefined.  There is no single preferred value for the central
surface brightnesses of disks in galaxies. There is only an upper limit
to the central surface brightnesses of disks, while for lower central
surface brightnesses the number of galaxies per volume element
decreases only slowly as function of the central surface brightness.\\
2)\ \ The Hubble sequence type index correlates strongly with the
effective surface brightness of the bulge, much better than with the
bulge-to-disk ratio.\\ 3)\ \ The disk and bulge scalelengths are
correlated.\\ 4)\ \ These scalelengths are not correlated with Hubble
type.  Hubble type is a lengthscale-free parameter and each type
therefore comes in a range of magnitudes (and presumably a range of
total masses).\\ 5)\ \ Low surface brightness spiral galaxies are not a
separate class of galaxies. In a number of aspects they are a
continuation of a trend defined by the high surface brightness
galaxies. Low surface brightness galaxies are in general of late Hubble
type.\\ 

\keywords{Galaxies: fundamental parameters -- Galaxies: luminosity function --
 Galaxies: photometry -- Galaxies: spiral -- Galaxies: statistics -- 
Galaxies: structure
 }
}

\begin{document}

 \thesaurus{03(11.06.2; 11.12.2; 11.16.1; 11.19.2; 11.19.7; 11.19.6)}
 \title{
Near-infrared and optical broadband surface photometry of 86 face-on
disk dominated galaxies. 
 \thanks{Based on observations with the Jacobus Kapteyn Telescope and the
Isaac Newton Telescope operated by the Royal Greenwich Observatory at
the Observatorio del Roque de los Muchachos of the Instituto de
Astrof\'\i sica de Canarias with financial support from the PPARC (UK) and NWO
(NL) and with the UK Infrared Telescope at Mauna Kea operated by the
Royal Observatory Edinburgh with financial support of the PPARC.\newline
% \indent Tables \ref{help} are also available in electronic form: see
%the Editorial in A\&A 1994 Vol.~280(3) p.~E1.
 }
} 
 \subtitle{III. The statistics of the disk and bulge parameters.}
 \author{Roelof S.~de Jong}
 \offprints{R.S.~de Jong, University of Durham, Dept.~of
Physics, South Road, Durham, DH1 3LE, United Kingdom, e-mail:
R.S.deJong@Durham.ac.uk} 
 \institute {Kapteyn Astronomical Institute, P.O.Box 800, NL-9700 AV
Groningen, The Netherlands}
 \date{received May 19, accepted Nov 5 1995}
 \maketitle
 \markboth{R.S.~de Jong}{Near-IR and optical broadband observations of 
86 spirals. III Statistics of bulge and disk}

%\begin{abstract}
%\abst
%\end{abstract}

\section{Introduction}
\label{Intro3}
 The light of a spiral galaxy is dominated by two components, the disk
and the bulge.  The basic difference between these components lies in
their support against gravitational collapse.  The disk is almost
completely rotationally supported, while the bulge is for some fraction
also pressure supported.  At least two parameters are needed to describe
the light distribution of each of these components: a surface brightness
term and a spatial scaling factor.  The fundamental parameters of the
disk are usually expressed in central surface brightness (\muo) and
scalelength ($h$), while the bulge parameters are expressed in effective
surface brightness (\mue) and effective radius (\re).  These fundamental
parameters were determined for a large statistically complete sample of
galaxies by de Jong (\cite{deJ2}, Paper~II).  The distributions of the
fundamental parameters are still poorly known and their statistics are
investigated in this paper with some emphasis on three relationships: 1)
``Freeman's law'', the empirical relation found by Freeman (\cite{Freeman})
indicating the
 %\footnote 
% {Freeman's law is in fact not a law, but an empirical relation found 
%by Freeman (\cite{Freeman}).  It has become known in the literature 
%as a law and therefore I will refer to it as such.}
 constancy of disk central surface brightness among galaxies, 2) the
number density of galaxies as a function of their fundamental parameters
and 3) the relation between the fundamental parameters and Hubble
classification. 

\subsection{Freeman's law}
 One of the most remarkable results presented in the classical paper of
Freeman (\cite{Freeman}) was the apparent constancy of the $B$ passband
\muo\ of spiral galaxies.  For a subsample of 28 (out of 36) galaxies
he found $\langle \mu_0\rangle \!= \!21.65 \pm 0.3$ $B$-mag arcsec$^{-2}$. 
If the central $M/L$ is approximately constant among galaxies, 
 %??as suggested by optical rotation curves (ref)??, 
 this translates directly into a constant central surface density of
matter associated with the luminous material. 

Several authors have tried to explain this result.  It has been argued
that ignoring the contribution of the bulge to the light profile could
produce the effect (Kormendy~\cite{Kor77}; Phillipps \&
Disney~\cite{PhiDis83}; Davies~\cite{Dav90}).  Freeman (\cite{Freeman})
did not decompose the luminosity profiles in a bulge and disk, but
fitted a line to the linear part of the luminosity profile plotted on a
magnitude scale.  This linear part of the profile could be contaminated
by bulge light.  With their models Kormendy~(\cite{Kor77}) and
Davies~(\cite{Dav90}) show that the central surface brightness of low
surface brightness disks will be overestimated by this procedure
because of the extra bulge light near the center.  The central surface
brightnesses of high surface brightness disks with a short scalelength
are underestimated; because of the small disk scalelength the bulge
light dominates the luminosity profile again in the outer region, but
with a longer scalelength and a lower surface brightness than the disk.
\mbox{Several} arguments can be raised against this interpretation (see also
Freeman~\cite{Fre78}):
 1) even with bulge light included the result is still important,
 2) many later type galaxies hardly have a bulge, but the
effect is still present (van der Kruit~\cite{Kru87}),
 3) in samples where proper decomposition techniques are used the
effect is still found, although with a larger dispersion
(Boroson~\cite{Bor81}),
 4) a limited range in bulge parameter space was explored in the models
mentioned above, which might not be representative of the bulges in spiral
galaxies. 

Dust extinction has also been proposed as an explanation for the
constancy of \muo\ (Jura~\cite{Jura}; Valentijn~\cite{Val90}).  If
galaxies are optically thick in the $B$ passband, one is only looking
one optical depth into the galaxies and always observes the same outer
layer.  This removes the inclination dependence from the Freeman
relation, but leaves the unsolved problem of why all galaxies should
have the same surface brightness at optical depth equal to one. 

Freeman established his relation in the $B$ passband where the light of
galaxies is dominated by a very young population of stars, which make
up only a few percent of the stellar mass.  Of all commonly used
passbands the light of the massive old stellar population is relatively
the most important in the near-infrared (near-IR) $K$ passband used
here. The $K$ passband has the additional advantage that the extinction
by dust is strongly reduced. The $K$ passband is therefore best suited
to trace the fundamental parameters of the luminous mass. However,
other passbands have been used as well in this study to investigate the
wavelength dependence of the bulge and disk parameters due to dust and
population effects. 

De Vaucouleurs (\cite{deV74}) was one of the first to suggest that the
constancy of \muo\ might result from a selection effect.  This was
later quantified by Disney (\cite{Dis76}) and Allen \& Shu
(\cite{AllShu79}).  Catalogs of galaxies have usually been selected by
eye from photographic plates using some kind of diameter limit.   One
might therefore select against very compact galaxies with a high
central surface brightness, because these have small isophotal
diameters.  Likewise, galaxies with a very low surface brightness might
have been missed due to the lack of contrast with the sky background. 
Disney \& Phillipps (\cite{DisPhi83}; see also Davies~\cite{Dav90})
define a visibility for a galaxy, which enables one to correct a sample
for these selection effects if one has made a careful initial sample
selection.

\subsection{Bivariate distributions}

Correcting for selection effects is in fact trying to determine from the
observed statistics how many galaxies there are per unit volume with a
certain property.  More than one property can be used in determining
such a distribution per volume.  One needs at least two parameters to
characterize the exponential light profile of a disk dominated galaxy
and a bivariate distribution function of both disk parameters is a more
general statistical description of galaxy properties than a one
parameter function.  The diameter, the central surface brightness and
the luminosity distribution functions of galaxies are integrations of
this bivariate distribution in a certain direction.  In this process
information is lost and the bivariate distribution function is therefore
more useful in studies of deep galaxy counts and provides more
constraints on theories of galaxy formation and evolution than its one
dimensional counterparts. 

Bivariate distribution functions of galaxies have been determined only
a few times before (Choloniewski \cite{Cho85};
 %determined the bivariate distribution of E and S0 galaxies, 
 Phillips \& Disney \cite{PhiDis86};
 %studied the distribution of a sample of Virgo spiral galaxies in the total
 %luminosity, central surface brightness plane and 
 van der Kruit \cite{Kru87}, \cite{Kru89};
 %) investigated the distribution of spiral
 %galaxies as function of central surface brightness and scalelength.
 Saunders et al.~\cite{Sau90};
 % presented the bivariate distribution of
 %60~$\mu$m flux versus isophotal diameter and magnitude. 
Sodr\'e \& Lahav \cite{SodLah93}).
 % calculated the diameter, magnitude bivariate function.   
 Even though different fundamental parameters are used, almost all
(except Saunders et al.) of these distributions describe fundamentally
the same thing in different ways.  These studies were performed in the
$B$ or comparable passbands, which is, as mentioned before, not the
wavelength most suited to study global fundamental properties of
galaxies.

\subsection{Morphological classification} For classification of spiral
galaxies on the Hubble sequence three principal discriminators are
used: 1) the pitch-angle of the spiral arms, 2) the degree of
resolution of the arms (into \hii\ regions, dust lanes and resolved
stars) and 3) the bulge-to-disk (B/D) ratio.  In his detailed
description of the Hubble sequence, Sandage (\cite{San61}) indicates
that the B/D ratio is the weakest discriminator unless galaxies are
seen edge-on.  He finds clear mismatches in type between
classifications using items 1) and 2) and classifications using item
3).  Another factor hampers the use of B/D ratio for classification of
early spirals.   On the photographs used for classification the central
region of an early spiral galaxy is normally overexposed in order to
show clearly the faint spiral structure. 

Still, the B/D ratio is often assumed to be the principle parameter
underlying the Hubble sequence, even though a tight correlation between
classification and measured B/D ratios was never found.  The
measurements indicate at best a trend (e.g.\ Simien \& de
Vaucouleurs~\cite{SimdeV86}; Andredakis \& Sanders~\cite{AndSan94}) and
the discrepancies between B/D ratio and Hubble type have been
attributed to two sources of error.  First there is the uncertainty in
classification.
 %??(the measurement of B/D ratio is right, but the classifications are wrong)??.  
 Comparisons of Hubble types given by different classifiers show
an rms uncertainty in type index of order 2 T-units (Lahav et al.~\cite{Lah95}).
The second source of error is the uncertainty in the bulge/disk
decomposition, 
 % ??(classification right, but B/D ratio wrong)??. 
 due to, among other things, the mathematical peculiarities of the
widely used \rq bulge law (de Vaucouleurs~\cite{deV48}).

\subsection{Outline}
 The main goal of this investigation is to determine the nature of the
Freeman law. In order to address the problems concerning the Freeman
law, a large sample of face-on spiral galaxies was carefully selected
and surface photometry was obtained in the $K$ passband as well as in
several other passbands. A large number of other global and structural
parameters of the galaxies were determined in this investigation and
their nature is also explored in this paper.

The remainder of this article is organized as follows. The data set and
the extraction of the observed bulge and disk parameters are briefly
described in Sect.~\ref{data3}.  The corrections to the observations
in order to calculate number distributions are described in
Sect.~\ref{corrections} and these distributions are presented for the
$B$ and the $K$ passband in Sect.~\ref{distr}.  The relations found
are discussed within the context of the three main points of interest
(Freeman's law, bivariate distributions and Hubble sequence) in
Sect.~\ref{discus3}. The conclusions are summarized in
Sect.~\ref{concl3}.

\section{The data}
\label{data3}
 In order to examine the parameters describing the global structure of
spiral galaxies, 86 systems were observed in the $B, V, R, I, H$ and $K$
passbands.  A full description of the selection, observations and data
extraction can be found in de Jong \& van der Kruit (\cite{deJ1},
Paper~I).  The galaxies in this statistically complete sample of
undisturbed spirals were selected from the UGC (Nilson~\cite{Nilson})
to have red diameters of at least two arcmin and minor over major axis
ratios larger than 0.625. The survey was limited to 12.5\% of the sky
globe. Standard reduction techniques were used to produce calibrated
images.

In Paper~II the extraction of the bulge and disk parameters from the
calibrated images is described.  An extensive error analysis was
performed using different fit techniques.  The best results were
obtained with a model galaxy with an exponential radial light profile
for both bulge and disk, that was two-dimensionally (2D) fitted to the
full calibrated image. This 2D fit technique made it also possible to
fit an additional Freeman bar (Freeman~\cite{Fre66}) component, which
improved the fit for 23 of the 86 galaxies.  The error analysis
revealed the two dominant sources of error in the derived component
parameters to be: 1) the assumed luminosity profile of the bulge and 2) the
uncertainty in the sky background subtraction. Other uncertainties, like
errors in seeing correction, zero-point errors and resolution problems
were found to be much smaller in most cases.

Assuming that the exponential profile is a reasonable description of
the bulge light distribution, the dominant source of error in the
parameters is caused by the uncertainty in the sky background level.
This uncertainty was not taken into account using the 2D fit technique,
but the 1D errors can be used, because the 2D fit results are generally
comparable to the double exponential 1D fit results (see Paper~II,
Fig.6).  The 1D errors do include the uncertainty of sky background
subtraction and are always larger than the formal 2D fit errors.
The 1D errors are only shown in the graphs presented here if they are
significantly larger than the symbol size. 

In this paper the RC3 (de Vaucouleurs et al.~\cite{rc3}) morphological
type index T is used (see also Paper~I). Because a few galaxies had no RC3
classification, I classified them as UGC\,1551--(8), UGC\,1577--(4),
UGC\,9024--(8) and UGC\,10437--(7).  The mean error in type index in
the RC3 is stated to be 0.89.  This number seems to be very low. Lahav
et al.~(\cite{Lah95}) showed that the dispersion between the RC3
T-index and the T-values of six expert classifiers was on average 2.2
T-units for a sample of 831 galaxies. The dispersion between any two
classifiers ranged between 1.3 and 2.3 T-units, with 1.8 on average. It
is safe to say that the uncertainty in classification in the RC3 is at
least 1.5 T-units.

The data set comprises 86 galaxies in six passbands.  To keep a clear
view on the obtained results I will concentrate on the two most extreme
cases, the $B$ and the $K$ passband data.  The results for the other
passbands are available in electronic form.  The $B$ and $K$ passband
results are displayed in the graphs with the same dynamic range (but
often with different zero-points) and therefore they can be compared
directly. 

\section{Corrections}
\label{corrections}

The observed bulge and disk parameters determined in Paper~II have to
be corrected for all kinds of systematic effects. These corrections
are often uncertain but necessary. One can only expect that they are at
least in a statistical sense correct. 

\subsection{Galactic foreground extinction}

The measurements of brightness and surface brightness were corrected
(unless stated otherwise) for Galactic foreground extinction according
to the precepts of Burstein \& Heiles (\cite{BurHei84}) and the actual
$B$ passband extinction values were adopted from the RC3.  The Galactic
extinction curve of Rieke \& Lebofsky (\cite{RieLeb85}) was used to
convert these $B$ passband extinction values to other passbands.  The
sample galaxies were selected to have a Galactic latitude larger than
25\degr; the extinction correction is in general small and gets smaller
for the longer wavelength passbands.  The average correction is 0.14
$B$-mag and the largest correction is 0.68 $B$-mag, which translates
into 0.06 $K$-mag. 
 %??There are other methods 
%for correcting for Galactic extinction (ref), which don't always 
%agree very well with Burstein \& Heiles method.  Their method is most 
%widely accepted and makes comparisons of our work against others 
%therefore easier.  
% All (surface) brightness distributions in this article have been
%corrected for Galactic extinction, unless stated otherwise.

\subsection{Inclination corrections}

Since Valentijn (\cite{Val90}) reopened the debate of optically thin
versus optically thick spiral galaxies, inclination corrections for
surface brightness have become less trivial.  A simple equation for
correcting surface brightnesses for inclination effects, taking
internal extinction into account, has the form
 \begin{equation}
\mu^i = \mu-2.5 C \log(a/b),
\label{inccor}
 \end{equation}
 where $a/b$ is the major over minor axis ratio of the galaxy and $C$
the internal extinction parameter, which takes values $0\!\le \!C\!\le
\!1$.  Fully transparent galaxies are described by $C\!=\!1$, while the
case $C\!=\!0$ describes the optically thick ones. 
 % Other $C$ values describe semi-transparent galaxies.  

It is unlikely that the inclination correction indeed takes such a form
in the optical passbands, as extinction in the optical passbands is for
a considerable fraction caused by scattering and not just by absorption
alone.  Light will be scattered preferably from edge-on directions to
face-on directions, which means that extinctions will seem to be higher
for edge-on than for face-on galaxies.  On top of that, certain
configurations of dust and stars can behave optically thin in an
inclination test, while they may in fact be completely opaque.  A clear
example of this is a very thin layer of optically thick dust between a
thicker slab of stars.  It is not trivial to produce a better
description as there are too many unknowns and $C$ itself may be a
function of galactic radius (see e.g.  Giovanelli et al.~\cite{Gio94};
Byun et al.~\cite{Byun94}). Therefore Eq.~(\ref{inccor}) is used as a
working hypothesis.  However, for a face-on selected sample such
corrections are small.  The average correction for the sample
examined here is 0.26 \magarc\ when $C\!=\!1$, with a maximum of 0.60
\magarc\ for the galaxy with the largest observed $b/a$ of 0.58. 

\subsection{Distances}

\begin{figure}
 \mbox{\epsfxsize=8.6cm\boundboxo{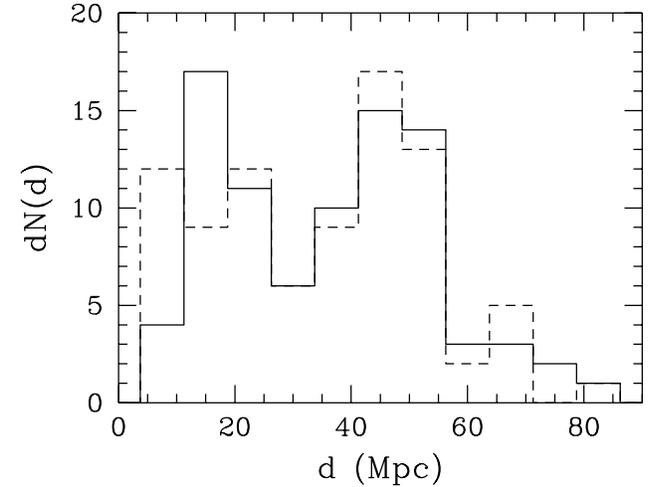}}
 \caption[]{
The distance distribution of the sample galaxies. For the dashed line
the $V_{\rm GSR}$ velocities from the RC3 were used, the full line
indicates the distance distribution when the velocities are corrected
for Virgo-centric infall. 
}
 \label{hisDist}
 \end{figure}

The distances to the observed galaxies were calculated using a Hubble
flow with an $H_0$ of 100 km s$^{-1}$ Mpc$^{-1}$, corrected for infall
into the Virgo cluster using the 220 model of Kraan-Korteweg
(\cite{Kra86}). This model assumes that the Local Group has an infall
velocity of 220 km/s towards the Virgo cluster and describes the
motions of the galaxies around the cluster by a non-linear flow model.
The $V_{\rm hel}$ velocities needed for this model were calculated from
the $V_{\rm GSR}$ velocities listed in the RC3, which are also
tabulated in Paper~I.  The nearest galaxy is at 6.2 Mpc, the most
distant galaxy is at 82.5~Mpc. The peculiar velocities of galaxies were
assumed to be on average 200 km/s in the line of sight, which
introduces an uncertainty ($\sigma_d$) of 2 Mpc in the distance estimates.
The distribution of distances is displayed in Fig.~\ref{hisDist}, which
shows some excess of galaxies at $\sim$45 Mpc because of an
extension of the Pisces-Perseus supercluster. The relationships
presented in this study are very little affected when other infall and
flow models are used to calculate distances. 

% and they will only be used in the
%parameterizations of distributions. 

\subsection{Selection correction}

The physically relevant quantities are not the observed numbers of
galaxies with a certain property, but the frequency of galaxies with a
certain property in a volume.  Therefore, the fact that a galaxy is
included in the sample has to be linked to the statistical probability
of finding such a galaxy in a certain volume.  The galaxies in the
sample were selected to have UGC red major axis diameter ($D_{\rm
maj}^{\rm lim}$) of at least 2 arcmin.  This creates a selection bias
against galaxies with low surface brightness and/or small scalelengths,
as they appear smaller on photographic plates.  The distances ($d$) to
the observed galaxies and their angular diameters ($D_{\rm maj}$) are
known and the maximum distance at which galaxy can be placed, while
still obeying the selection criteria, can be calculated ($d_{\rm max}
\!= \! dD_{\rm maj}/D_{\rm maj}^{\rm lim}$).  A galaxy can only enter
the sample if it lies in a spherical volume which has this maximum
observable distance as radius.  Turning this argument around, one can
expect on statistical grounds that a selected galaxy samples a spherical
volume with a radius equal to its maximum observable distance (a more
formal discussion can be found in Felten~\cite{Fel76}).  The volume
sampled by a galaxy in a diameter limited sample is thus
 \begin{equation}
V_{\rm max} = \frac{4\pi}{3}(d_{\rm max})^3 = \frac{4\pi}{3}(dD_{\rm maj}/D_{\rm maj}^{\rm lim})^3.
\label{vmax}
 \end{equation}
 Following the previous line of reasoning, an estimate for the average
number of galaxies in a unit volume obeying a certain specification
($S$) {\em for a complete sample of $N$ galaxies} is
 \begin{equation} 
 \Phi(S) = \sum_{i}^{N} S^{i}/V^{i}_{\rm max} ,
 \label{vmaxcor}
 \end{equation} 
 where $i$ is summed over all $N$ galaxies in the sample and
$S^{i}\!=\!1$ if the specification is true for galaxy $i$ and
$S^{i}\!=\!0$ if false.  
 %%?? Think about using $\Phi(S)\,ds$??
 The error in $\Phi(S)$, assuming Poison statistics in a homogeneous
universe and considering the uncertainties in the distances, can be
calculated by
 \begin{equation} 
\sigma_{\Phi(S)}^2 = \sum_{i}^{N} (S^{i}/V^{i}_{\rm max})^2
+\sigma_d^2 \sum_{i}^{N} (3 S^{i}/d V^{i}_{\rm max})^2 .
 \end{equation} 

There is always a chance that a member of a peculiar class of galaxy
happens to be nearby and gets a lot weight in Eq.(\ref{vmaxcor}) and
this volume correction can therefore only be applied to large samples. 
One must ensure that a large enough volume of space is sampled so that
galaxies are randomly distributed in space.  Figure~\ref{hisDist} shows
that the sample mainly traces the local density enhancement, as large
scale structures in the universe have scales of order 50 Mpc. 
Equation~\ref{vmaxcor} should therefore be used with care, because the
number of galaxies with small intrinsic diameters will be overestimated
relative to the larger ones due to the local density enhancement.  The
average number of galaxies per Mpc$^3$ calculated with
Eq.~(\ref{vmaxcor}) might be more representative of the local
environment than of the mean cosmological values.  Still it is a useful
equation to observe general trends in bivariate distributions and to
compare results obtained from different passbands. 

Other methods to correct distributions for selection effects have been
advocated, because they take spatial density fluctuations into account
(for an overview see Efstathiou et al.~\cite{Est88}). These methods
assume that the intrinsic distribution function is independent of
position ({\bf x}) in space, so that we can write $\Phi(S) \!=
\!\phi(S)\rho({\bf x})$, thereby losing the absolute calibration of the
number density. These methods all assume a clear relation between the
distribution parameter(s) and the limiting selection parameter(s). This
is not the case for the current investigation. A diameter limit is not
trivially linked to the central surface brightness distribution,
certainly not when a different passband is used for the selection and the
distribution. 

The correction of Eq.~(\ref{vmaxcor}) is only valid if a particular
galaxy would have been measured at the same intrinsic (as opposed to
angular) diameter, had it been at a different distance.  In Paper~I it
was shown that this is probably the case for the UGC galaxies with type
index T$\le$6.  For later types the situation is less clear, there is a
too short a range in diameters to check and it must be assumed that for
late-type systems the same type of galaxy is measured at the same
intrinsic diameter at different distances.  Under this assumption it is
{\em not} important that the UGC eye estimated diameters of late-type
galaxies correspond to lower average surface brightness than that of
early types (see Paper~I, Fig.~11).  This effect just means that there
are more late-type galaxies in the sample than expected based on
their isophotal diameter, but their average distance will be larger so
that the number of galaxies per sampled volume stays the same. 

The volume correction of Eq.~(\ref{vmaxcor}) can be used to calculate
number density distributions for all passbands, as long as the red UGC
diameters are used to calculate the $V_{\rm max}$. The distribution of
any galaxy parameter $S^{i}$ can be determined in any passband; the
use of the red UGC diameters in Eq.~(\ref{vmaxcor}) ensures the
correction for the intrinsic selection effects of the whole sample.  

Next to the diameter limit, there are two more selection criteria
defining the sample.  The selection was limited to 12.5\% of the sky and
only galaxies with $b/a\!>\!0.625$ were used, which is only 37.5\% of all
possible random orientations.  Equation~\ref{vmaxcor} was corrected for these
 %additional
 selection criteria.  A correction was also applied for the
fraction of galaxies for which no (photometric) data was available in a
certain passband.  All these corrections were made under the assumption
that the incompleteness had no correlation with the investigated
parameters. 

%??In appendix:??

Equation~(\ref{vmaxcor}) can only be applied when the sample is
complete.  The statistical completeness of the sample can be tested with
the $V/V_{\rm max}$-test (Paper~I).  The $V/V_{\rm max}$ of a galaxy is
the spherical volume associated with the distance of a galaxy divided by
$V_{\rm max}$ as defined in Eq.(\ref{vmaxcor}), thus for a galaxy in
this diameter limited sample $V/V_{\rm max} \!= \!(D_{\rm maj}^{\rm
lim}/D_{\rm maj})^3$.  For objects distributed randomly in space the
average value of $V/V_{\rm max}$ should be $0.5 \pm 1/\sqrt{12\times
N}$, where $N$ is the number of objects in the test.  For the current
sample $\langle V/V_{\rm max}\rangle \!= \!0.57 \pm 0.03$ and therefore
there are slightly too many galaxies with a small angular diameter in the
sample.  The original sample of 368 galaxies from which the current
subsample was selected had a $\langle V/V_{\rm max}\rangle \!= \!0.496
\pm 0.015$ (Paper~I).  Subsequent selection depended only on the
position on the sky and therefore the excess of small diameter galaxies
is probably caused by the density enhancement of the Pisces-Perseus
supercluster, which gives some extra galaxies at the diameter selection
limit.  This might give some extra high surface brightness and/or large
scalelength galaxies in the sample above the cosmological mean, because
galaxies have to be intrinsically large to be included in the sample
being at the distance of the Pisces-Perseus supercluster. 

In a recent paper Davies et al.~(\cite{Dav94}) argued that the sample used
by van der Kruit (\cite{Kru87}) was incomplete in a magnitude $V/V_{\rm
max}$-test.  They argued that a hidden magnitude limit had influenced
the selection, so that an extra selection correction should be applied. 
I will follow up on this argument as the sample used here has been
selected using similar criteria as van der Kruit used for his sample. 
\footnote{Davies et al.~(\cite{Dav94}) also indicate that van der Kruit's
sample becomes incomplete for low surface brightnesses at $\muo\!>\!22.3$ as
$\langle V/V_{\rm max}\rangle \!= \!0.35 \pm 0.08$.  I would like to note
that this might just be a statistical fluctuation of low number
statistics, as $\langle V/V_{\rm max}\rangle \!= \!0.41 \pm 0.10$ for
$\muo\!>\!22.5$ and $\langle V/V_{\rm max}\rangle \!= \!0.46 \pm 0.13$ for
$\muo\!>\!22.7$, and thus for even lower surface brightnesses the sample is in
the statistically complete range of $\langle V/V_{\rm max}\rangle \!= \!
0.5$.}

There is nothing hidden about a magnitude selection effect for a
diameter limited sample.  On the contrary, it is expected.  For galaxies
with a certain absolute magnitude $M$ there exists a range of possible
(\muo,$h$) combinations, satisfying $M \!\propto \!\muo\!-\!5\log(h)$,
but only a limited range of them will satisfy the diameter selection
criterion $D_{\rm maj} \!\propto \!(\mu_{\rm lim}\!-\!\muo)h \!>
\!D_{\rm maj}^{\rm lim}$.  Thus for galaxies of the same apparent
magnitude we will miss some of the small scalelength, bright \muo\ and
some of the large scalelength, faint \muo\ galaxies, while still having
selected a complete sample in diameter.  The complication arises because
apparent diameters and magnitudes are not independent parameters for
galaxies and their $V/V_{\rm max}$-tests cannot be applied
independently.  Similar to Eq.~(\ref{vmax}), a $V_{\rm max}$
corresponding to a magnitude limit can be constructed.  The smallest of
the magnitude $V_{\rm max}$ and the diameter $V_{\rm max}$ values should
be used for each galaxy in a combined $V/V_{\rm max}$-test.  These tests
cannot be performed separately.  The fundamental premise of a diameter
limited sample is that the diameter $V_{\rm max}$ is always smaller than
the magnitude $V_{\rm max}$, and therefore a ``hidden'' magnitude limit
does not have to be taken into account and Eq.~(\ref{vmaxcor}) is
sufficient. 

%??A simplified example is instructive to show the difference more clearly. 
%Suppose the universe was made of two types of galaxies, all having the
%same absolute magnitude, but one having twice the isophotal diameter of
%the other in kpc.  We would select the larger diameter galaxies up to
%twice the distance of the small diameter galaxies.  Assuming equal
%volume number densities, we would have eight times as much large diameter
%galaxies in our sample as small diameter ones.  A magnitude $V/V_{\rm
%max}$-test would indicate that our sample starts to get incomplete at the
%apparent magnitude associated with the distance where the small diameter
%galaxies no longer satisfy the apparent diameter selection criterion.  A
%diameter $V/V_{\rm max}$ would indicate that this sample is complete in
%diameter selection, and more important, the volume correction described
%by Eq.~(\ref{vmaxcor}) is perfectly capable to correct this sample for
%its selection effects.  We can conclude that even though a diameter
%selected sample is incomplete in a magnitude $V/V_{\rm max}$-test (and
%is supposed to be so), we do not need to make extra magnitude selection
%correction. ??
%\input{lapcs}
\begin{figure}[t]
 \mbox{\epsfxsize=8.6cm\boundboxo{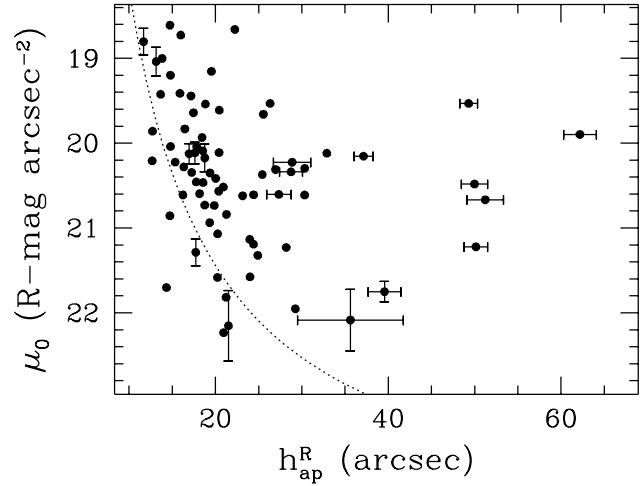}}
 \caption[]{
 The apparent scalelength versus observed central surface brightness of
the disks.  The dashed line indicates the selection limit of $D_{\rm
maj}\!\ge \!2\arcmin$ at the 24.7\,$R$-\magarc\ isophote for galaxies
with perfect exponential disks.  The indicated error estimates are the
errors from the 1D profile fitting with exponential bulge taking the
uncertainty in the sky background level into account. The errors are only
plotted if they are significantly larger than the symbol size. 
}
 \label{lapcs}
 \end{figure}

%?? Something about visibilities Davies, Allen and Shu, Phillips and
%Disney?? 

\section{The distribution of disk, bulge and bar parameters}
\label{distr}

In this section I investigate the distributions of the structural
parameters of the different galaxy components as a function of
morphological type and of each other. First the structural parameters
of the disk and bulge 
 %and bar 
 are examined independently. In the final
subsection, the relationships between disk and bulge parameters are
investigated. The distributions of bulge and disk parameters are
corrected for selection effects to yield volume number densities.

\subsection{The disk parameters}

Figure~\ref{lapcs} indicates some aspects of the completeness and
selection effects of the sample. It shows the distribution of the
observed central surface brightnesses versus scalelength as obtained
from the 2D fits of Paper~II.  The $R$ passband values are plotted,
because these values are most closely related to the (red UGC diameter)
selection criteria.  The dotted line indicates the selection limit for
a diameter cutoff at 2 arcmin at a surface brightness of 24.7
$R$-\magarc\ for a perfect exponential disk.  The 24.7~$R$-\magarc\ is
the average surface brightness at which the UGC red diameters were
determined (see Paper~I).  As mentioned in Paper~I, not all UGC
galaxies had their diameters estimated at the same isophote level. 
This explains why there are some galaxies to the left of the selection
line in Fig.~\ref{lapcs}.  If all galaxies have the same scalelength,
the number of galaxies expected in the sample will decrease as $h_{\rm
ap}^3$ and therefore it is not surprising that there are hardly any
galaxies in the sample below 22~$R$-\magarc. Obviously no galaxies can
enter the sample with \muo\ fainter than $\sim$24.7 $R$-\magarc. 

{\begin{table*}
\tabcolsep=1.9mm
\begin{tabular}{cc@{\ \ }rcccc@{\ \ \ }rccc}
\hline
\hline
         & & \multicolumn{4}{c}{$\langle \mu_0 \rangle$ ($B$-\magarc)} & & \multicolumn{4}{c}{$\langle \mu_0 \rangle$ ($K$-\magarc)} \\
RC3 type & & nr. & $C=0$ & $C=0.5$ & $C=1$& & nr. & $C=0$ & $C=0.5$ & $C=1$\\
\hline
\ $0\le T < \ 6$  & & 61 & 21.32 $\pm$ 0.78 & 21.45 $\pm$ 0.76 & 21.58 $\pm$ 0.74 & & 60 & 17.48 $\pm$ 0.71 & 17.61 $\pm$ 0.69 & 17.75 $\pm$ 0.67\\
\ $6\le T < \ 8$  & & 12 & 22.01 $\pm$ 0.75 & 22.16 $\pm$ 0.73 & 22.30 $\pm$ 0.72 & & 10 & 18.34 $\pm$ 0.90 & 18.50 $\pm$ 0.90 & 18.65 $\pm$ 0.90\\
\ $8\le T \le 10$ & &  8 & 22.97 $\pm$ 0.60 & 23.12 $\pm$ 0.57 & 23.26 $\pm$ 0.55 & &  7 & 20.05 $\pm$ 1.05 & 20.21 $\pm$ 1.03 & 20.37 $\pm$ 1.01\\
\ all             & & 81 & 21.59 $\pm$ 0.92 & 21.72 $\pm$ 0.90 & 21.86 $\pm$ 0.89 & & 77 & 17.82 $\pm$ 1.08 & 17.96 $\pm$ 1.08 & 18.10 $\pm$ 1.08\\
\hline
\hline
\end{tabular}
\caption[]{
 The average Galactic extinction corrected central surface brightnesses
for different inclination corrections (Eq.~(\ref{inccor})) and type index
bins.  $C=0$ corresponds to an optically thick disk, $C=0.5$ to a semi
transparent disk and $C=1$ to a fully transparent disk.  The values are
in \magarc\ with their standard deviations. 
 \label{avecsb}
}
\end{table*}
}

\begin{figure*}
 \mbox{\epsfxsize=8.8cm\boundboxo{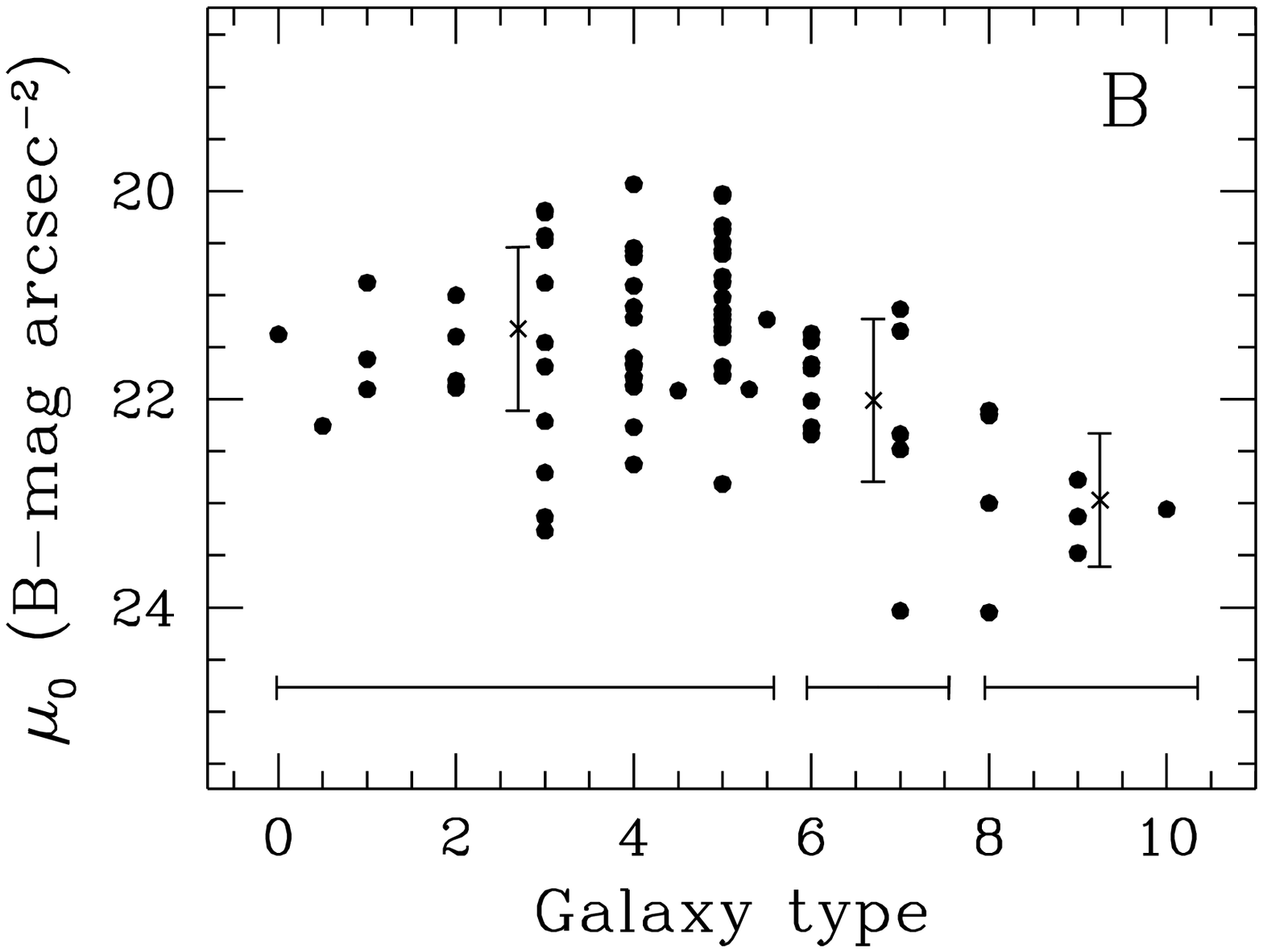}}
 \mbox{\epsfxsize=8.8cm\boundboxt{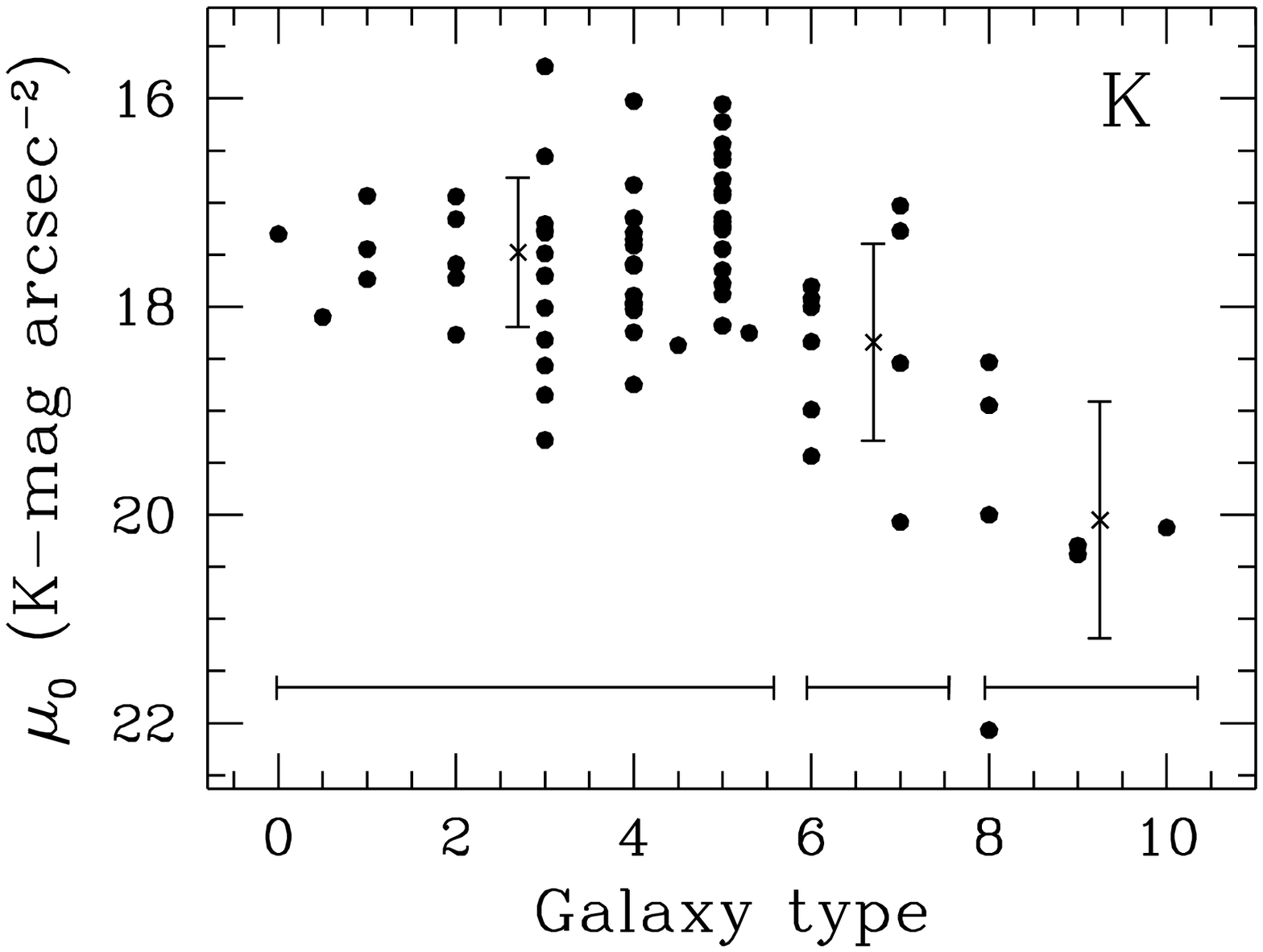}}
 \caption[]{
The Galactic extinction corrected central surface brightness of the
disks as function of morphological RC3 type. The crosses show the
values averaged over the bins indicated by the horizontal bars. The
vertical bars indicate the standard deviations of the mean values. 
\label{type_cs}
}
 \end{figure*}

\begin{figure*}
 \mbox{\epsfxsize=8.8cm\boundboxo{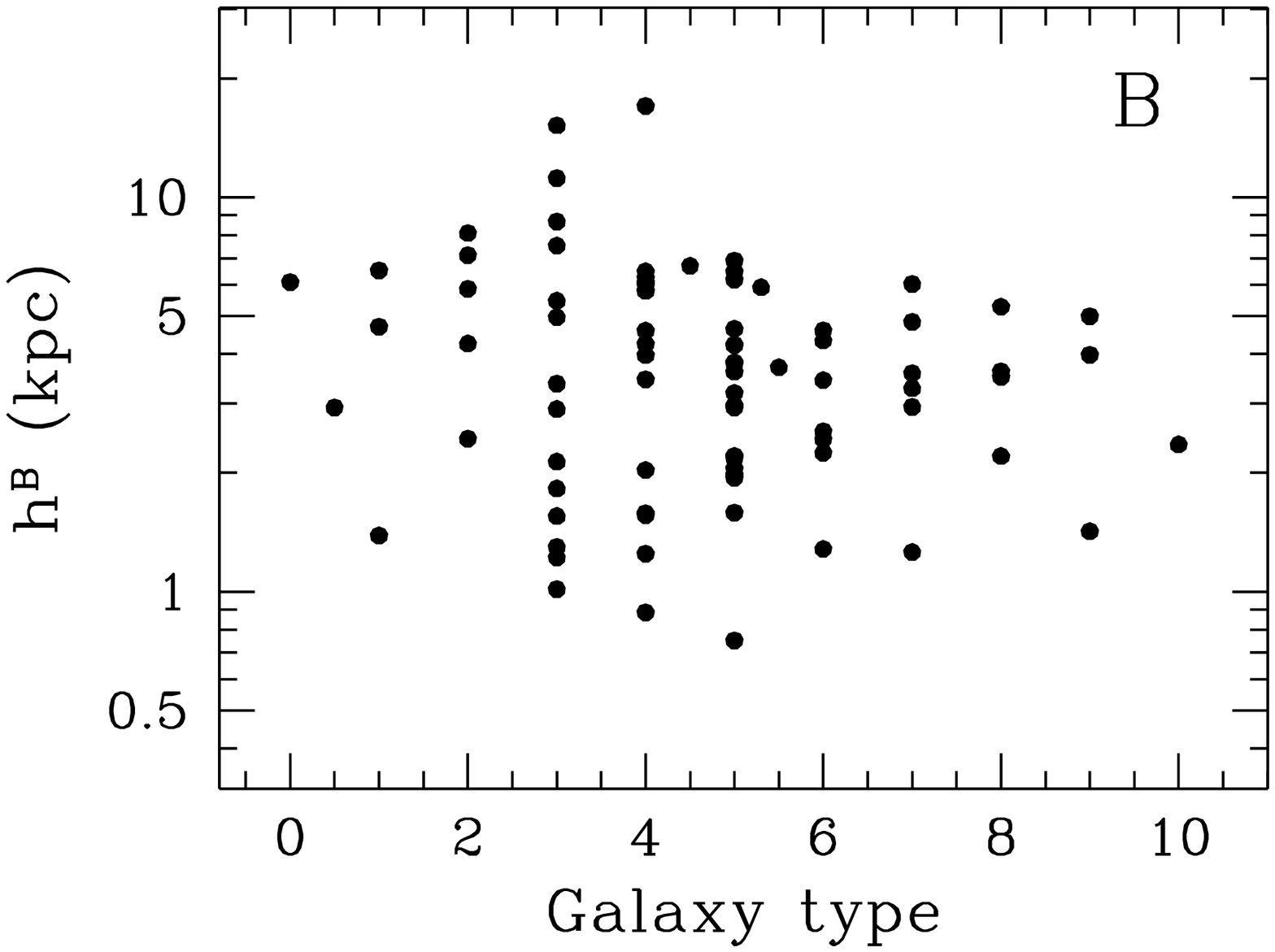}}
 \mbox{\epsfxsize=8.8cm\boundboxt{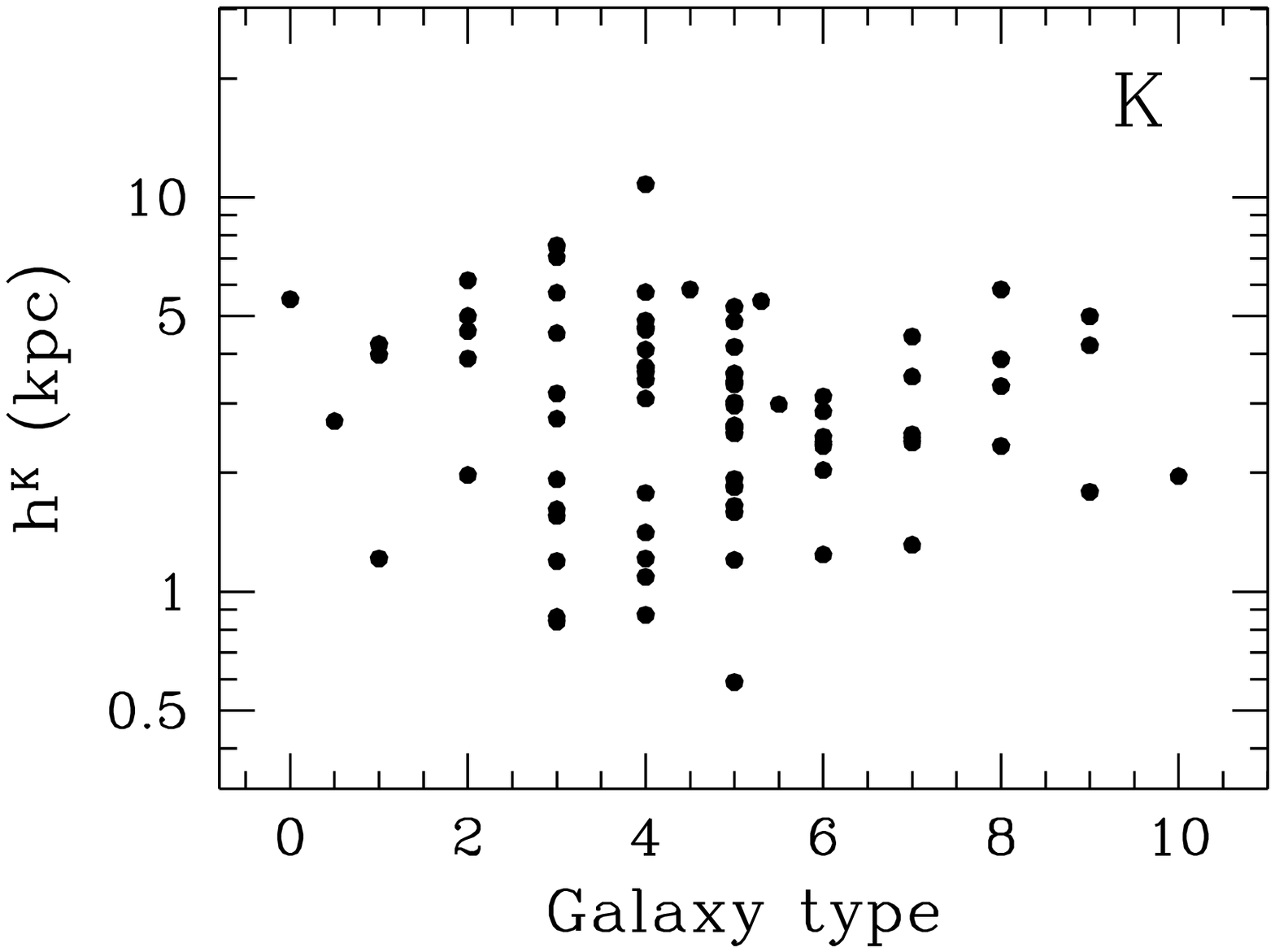}}
 \caption[]{
The scalelength of the disk as function of morphological type.
 \label{type_sc}
}

 \end{figure*}

Let us now look at the central surface brightness as function of
morphological type (Fig.~\ref{type_cs}).  Apparently the galaxies from
type T$\ \!= \!1$ to 6 have on average the same \muo, but with a large
scatter.  The later types have on average a significantly lower central
surface brightnesses and they might be classified as late types just
because they are low surface brightness (LSB) systems. 
 %We will see
%later that B/D ratio is not the main parameter for galaxy
%classification.  The original Hubble sub classification for spirals was
%based on the structure of the spiral arms.  On the POSS plates the LSB
%disks can hardly be seen. 
 It can be readily seen that this difference between early and late-type
galaxies increases when going from the $B$ to the $K$ passband.  This
indicates that disks of the later type spirals are bluer than the disks
of the early ones, but the discussion on the colors of these galaxies is
postponed to Paper~IV in this series (de Jong~\cite{deJ4}).  The average
\muo\ values were calculated for three morphological type bins indicated
by the horizontal bars in Fig.~\ref{type_cs} as well as for the total
sample.  The values with their standard deviations are tabulated in
Table~\ref{avecsb}. 

\begin{figure*}
 \mbox{\epsfxsize=8.8cm\boundboxo{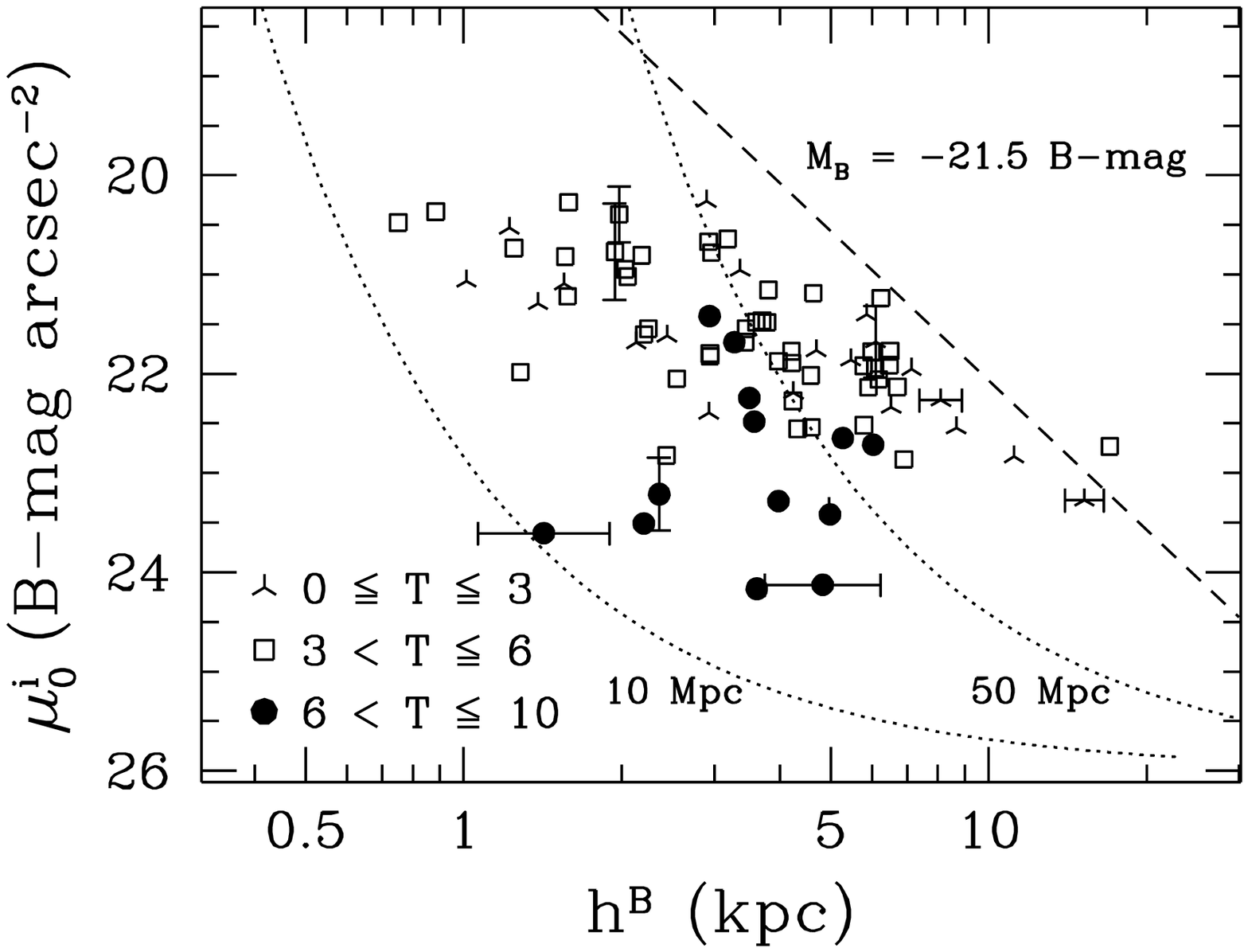}}
 \mbox{\epsfxsize=8.8cm\boundboxt{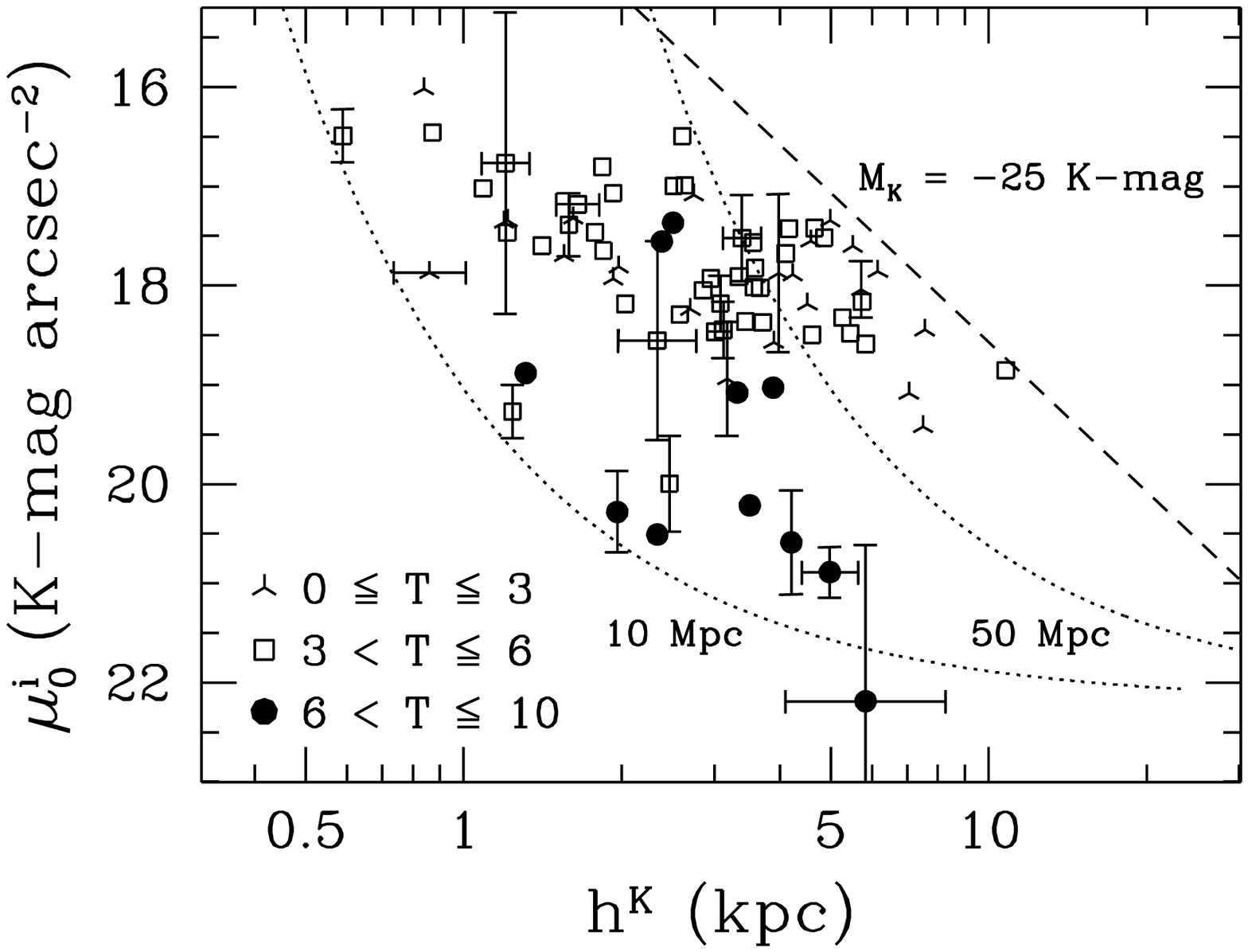}}
 \caption[]{
 The scalelength of the disks versus the central surface brightness. 
Different symbols are used to denote the indicated morphological type
ranges.  Exponential disks with equal absolute luminosity of indicated
magnitude are found on the dashed line.  Equality lines of other
magnitudes lie parallel to the dashed line.  The dotted lines indicate
the selection limits for all exponential disk galaxies closer than 10
and 50\,Mpc respectively, under the assumptions made in the text. 

%.  To the right of these lines we should be 
%complete to the indicated distance
 \label{sclcs}
}

 \end{figure*}

\begin{figure*}
 \mbox{\epsfxsize=8.8cm\boundboxo{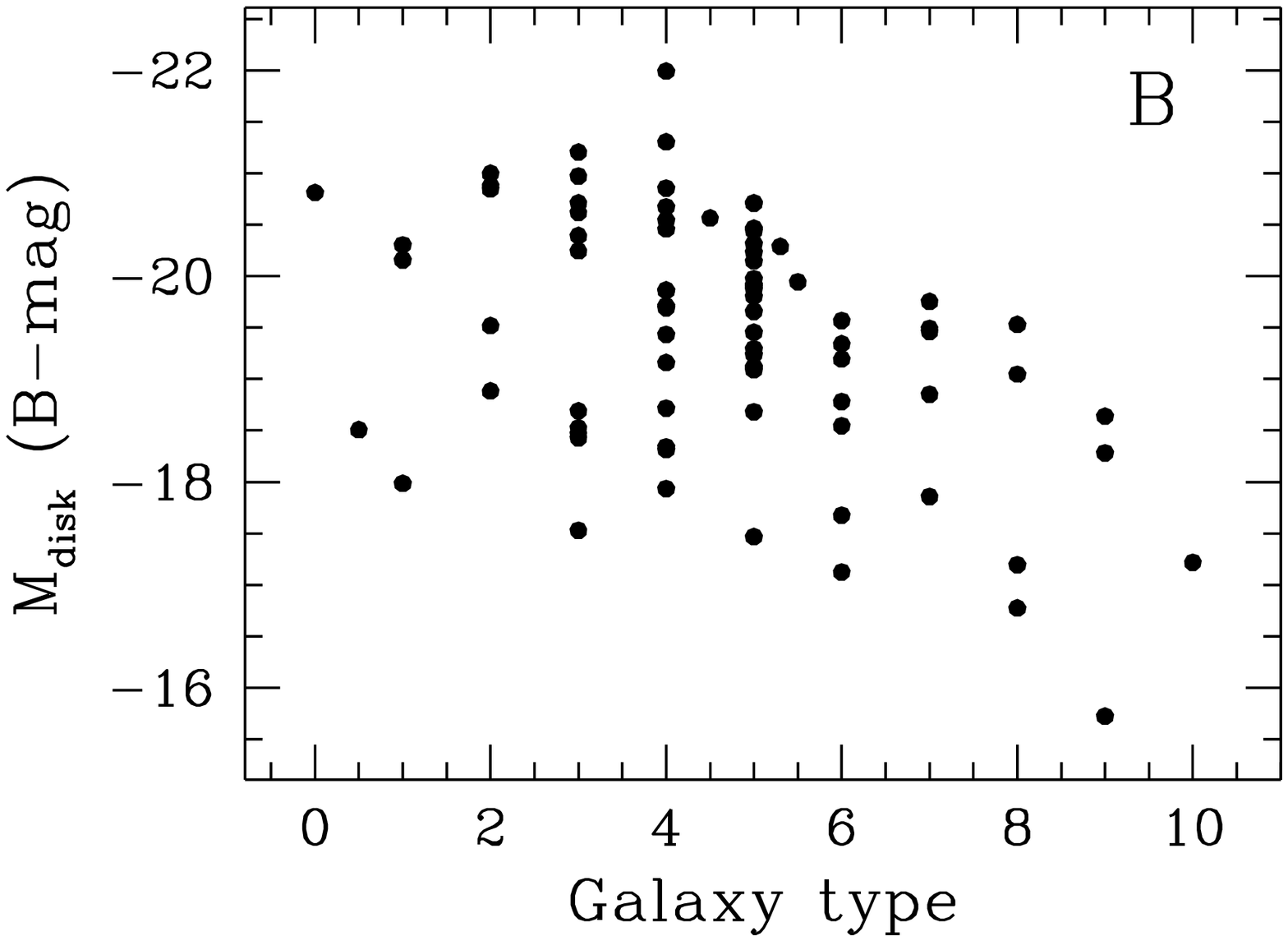}}
 \mbox{\epsfxsize=8.8cm\boundboxt{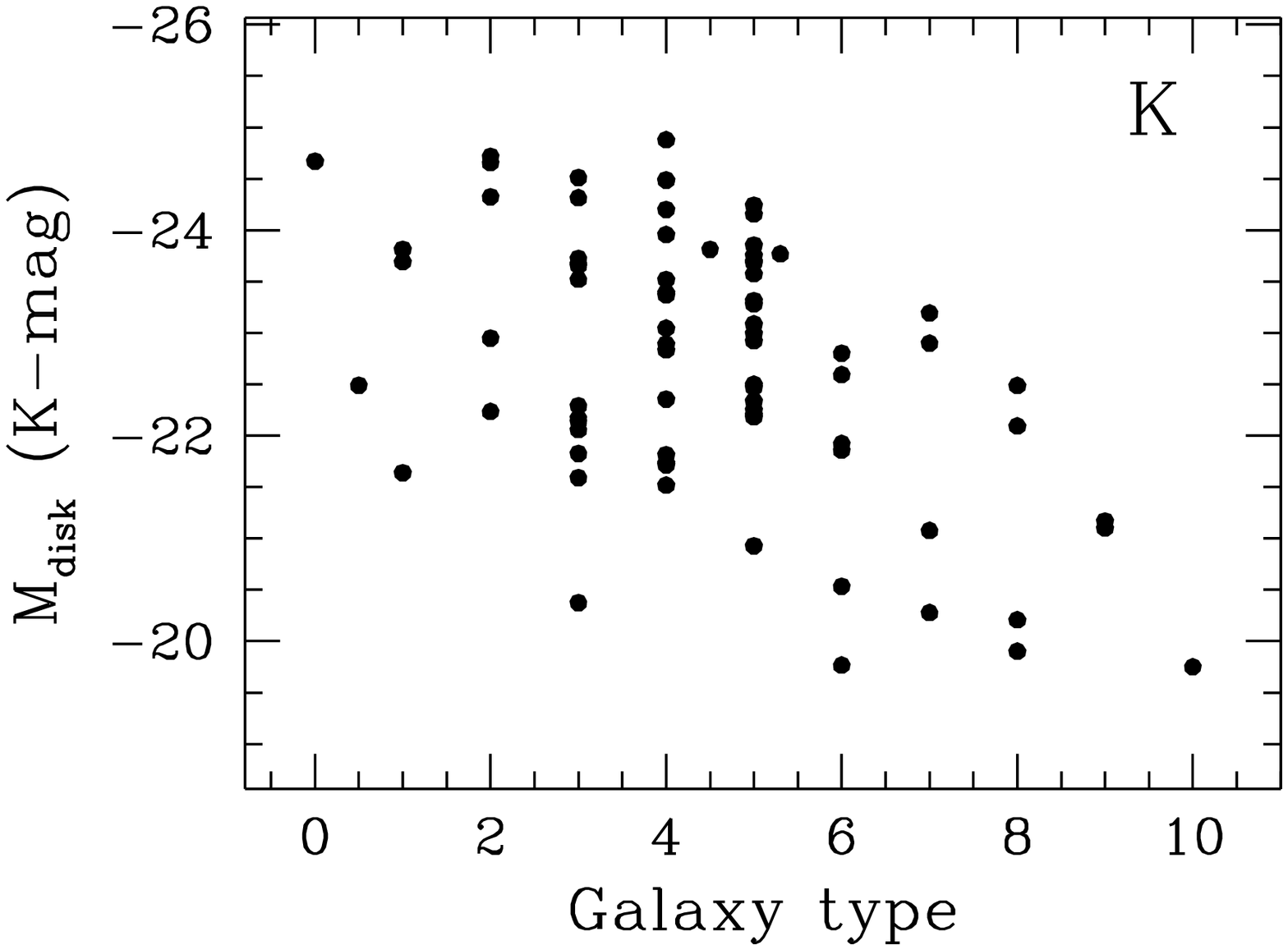}}
 \caption[]{
The distribution of absolute disk magnitudes as function of type index.
}
 \label{type_Md}
 \end{figure*}

The average \muo\ values were also calculated with an inclination
correction according to Eq.~(\ref{inccor}) with values for $C \!=\!0.5$
and $C\!=\!1$ (semi transparent and completely transparent behavior). 
The results can also be found in Table~\ref{avecsb}.  The standard
deviations on the average  \muo\ values are slightly smaller for
$C\!=\!1$, and even though it is a small effect, it is persistent for
all subgroups and all passbands.  The main result is of course a shift
in the mean central surface brightness of the disks.  For all remaining
plots an inclination correction with $C\!=\!1$ will be used. 

The distribution of the other disk parameter, the scalelength ($h$), as
function of type is shown in Fig.~\ref{type_sc}.  There is no trend of
$h$ with type and there is a large range in scalelengths.  There might
be a lack of late-type galaxies with small scalelengths, but this can
probably be attributed to a selection effect: the selection criteria
are heavily biased against LSB galaxies with small scalelengths.  The
scalelengths are smaller in the $B$ passband than in the $K$ passband
(discussion in Paper~IV).

The information of Figs~\ref{type_cs} and \ref{type_sc} are combined in
Fig.~\ref{sclcs}.  This figure shows that there is an upper limit in
the (\muo,$h$)-plane, as there are no galaxies with large scalelengths
and high central surface brightnesses.  This cannot be caused by
selection effects, large bright galaxies just cannot be missed in a
diameter selected sample.  This upper limit has been noted before by
Grosb\o l (\cite{Gro85}).  The upper limit partly follows the line of
constant total disk luminosity, as indicated by the dashed line in
Fig.~\ref{sclcs}.  Note that the Tully-Fisher relation
(\cite{TulFis77}, hereafter TF-relation) implies that this is also a
line of constant maximum rotation speed of the disk. There is also an
upper limit to the central surface brightness at about 20 $B$-\magarc\
(16 $K$-\magarc).  Again galaxies brighter than these limits are hard
to miss because of selection effects. 

% also line of constant angular momentum

%\input{bilogsccs}
\begin{figure*}
 \mbox{\epsfxsize=8.8cm\epsfbox[87 140 482 440]{pbf.cps}}
 \mbox{\epsfxsize=8.8cm\epsfbox[67 140 462 440]{pkf.cps}}
 \caption[]{
 The volume corrected bivariate distribution of galaxies in the
($\mu_0$,$h$)-plane.  The number density $\Phi(\muo^i$,$h$) is per bin
size, which is in steps of 0.3 in log($h$) and 1 \magarc\ in $\muo^i$. 
 }
 \label{bilogsccs}
 \end{figure*}

Late-type galaxies have lower central surface brightnesses in
Fig.~\ref{sclcs}, but the early and intermediate types show no
segregation.  The scalelengths also gives no segregation according to
type.  Very few late-type galaxies with very short scalelengths were
selected, but as shown before, late-type galaxies have lower surface
brightnesses and the selection biases against galaxies with low surface
brightness and short scalelengths are large. These biases are indicated
by the dotted lines in Fig.~\ref{sclcs}. To the right of these lines
the sample should be complete to the indicated distance. The lines are
calculated under the assumption that all galaxies have perfect
exponential disks with the same color at the selection radius
($B$--$R\! =\! 1.3$, $R\,$--$K \!= \!2.5$) and that the selection limit
is at 2\arcmin\ diameter at the 24.7 $R$-\magarc\ isophote (as in
Fig.~\ref{lapcs}).  Although these assumptions are not valid for an
individual galaxy, the dotted lines help to estimate the selection
effects; the galaxies near the 50 Mpc line had about 125 times more
chance of being included in the sample 
 %(are 125 times more
%``visible'' according to the definition of Phillipps \& Disney~\cite{PhiDis83})
 than the galaxies near the 10 Mpc line! The fact that the number
density of objects does not decrease by 125 from one line to the other
already indicates that there are many more ``small'' galaxies per
volume element than ``large'' galaxies.
 %(Remember that the sampled volume goes as the maximum selection 
%distance cubed.)

The distribution of the absolute magnitude of the disk ($M_{\rm disk}$)
against type (Fig.\ref{type_Md}) can also be deduced from
Figs~\ref{type_cs} and \ref{type_sc} ($M_{\rm disk}\! \propto\! \muo\!-\!2.5 \log (2
\pi h^2)$, no inclination dependent extinction correction was
applied).  As scalelengths show little correlation with type, the
distribution of disk magnitudes reflects the distribution of the central
surface brightness against type. There was no apparent segregation
according to bar classification in Figs~\ref{type_cs}, \ref{type_sc},
\ref{sclcs} and \ref{type_Md}.

% The bar classification had no influence on the distribution
%of \muo\, $h$ and $M_{\rm disk}$. 

%Also observed, uncorrected distribution of \muo

So far, only the observed distributions were presented, but the
distributions per volume are of more importance.  Therefore the volume
correction as described in Sect.~\ref{corrections} was applied. The
correction transforms Fig.~\ref{sclcs} into the bivariate distribution
in the (\muo,$h$)-plane presented in Fig.~\ref{bilogsccs}.  This is a
representation of the true number distribution of spiral galaxies per
volume element of one Mpc$^3$ with respect to both disk parameters. 
The magnitude and \muo\ upper limits noticed in Fig.~\ref{sclcs} are
also present here.  We are dealing with low number statistics now,
which is reflected in the erratic behavior of the distribution.  The
uncertainty increases in the direction of small scalelength and low
surface brightness.  These galaxies have so small isophotal diameters
that they really have to be nearby to be included in the sample and
such a small volume is sampled that statistics are working against us. 
For example if the true volume densities in the (17\,$K$-\magarc,
1\,kpc) and (21\,$K$-\magarc, 1\,kpc)-bins are equal, the chance of
observing a galaxy in the last bin would be 0.5.  If there had been
such a galaxy in the sample, a lot of weight would have been given to
it.  In short, the distributions are not well sampled in the low
surface brightness, small scalelength region.  No galaxies were
selected in this region, but the traced volume is also very small.  The
dominant type of spiral galaxy has a scalelength of about 1\,kpc and a
central surface brightness of 21~$B$-\magarc\ (17~$K$-\magarc). 

{
\begin{figure*}
 \mbox{\epsfxsize=8.8cm\boundboxo{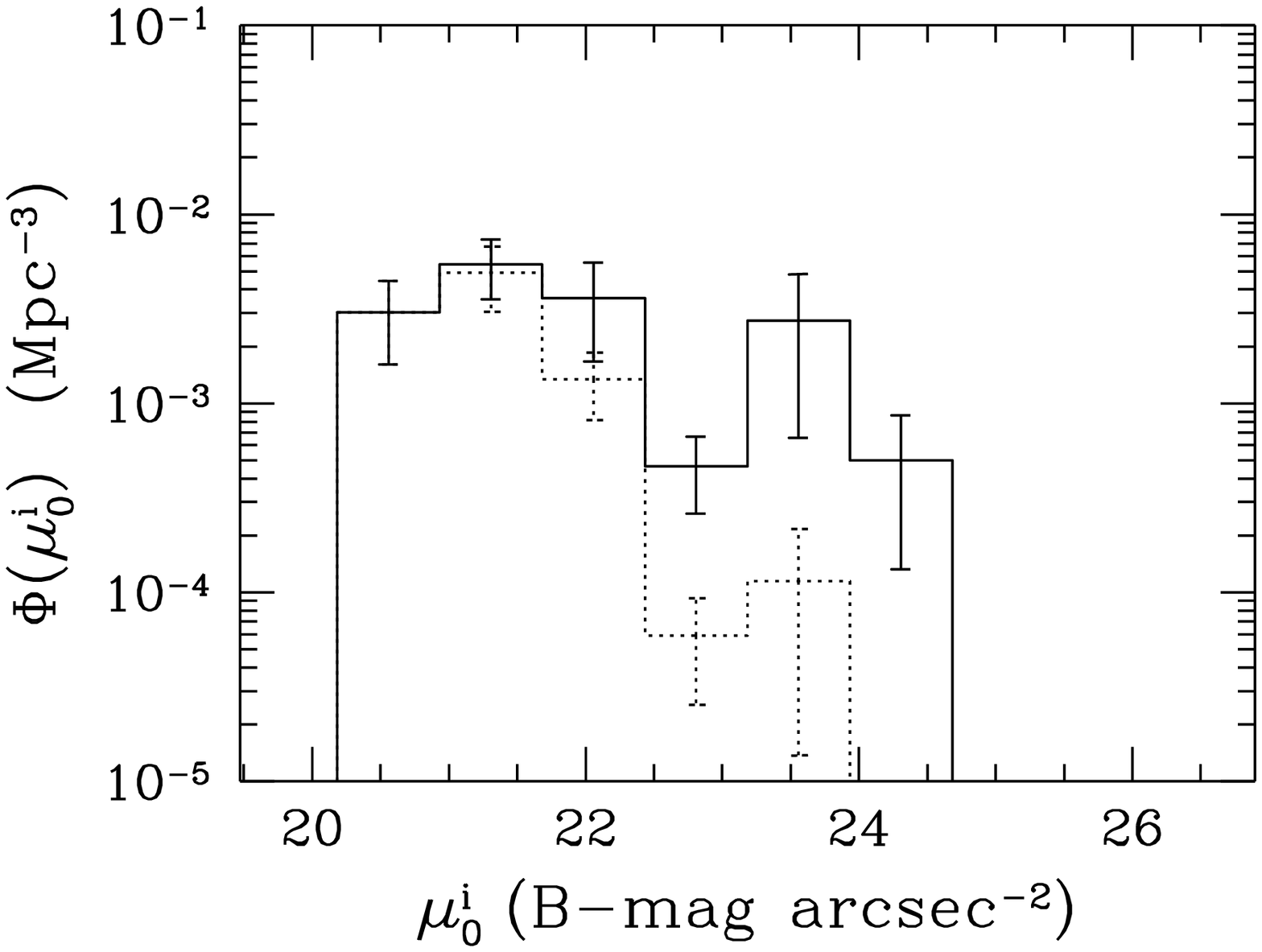}}
 \mbox{\epsfxsize=8.8cm\boundboxt{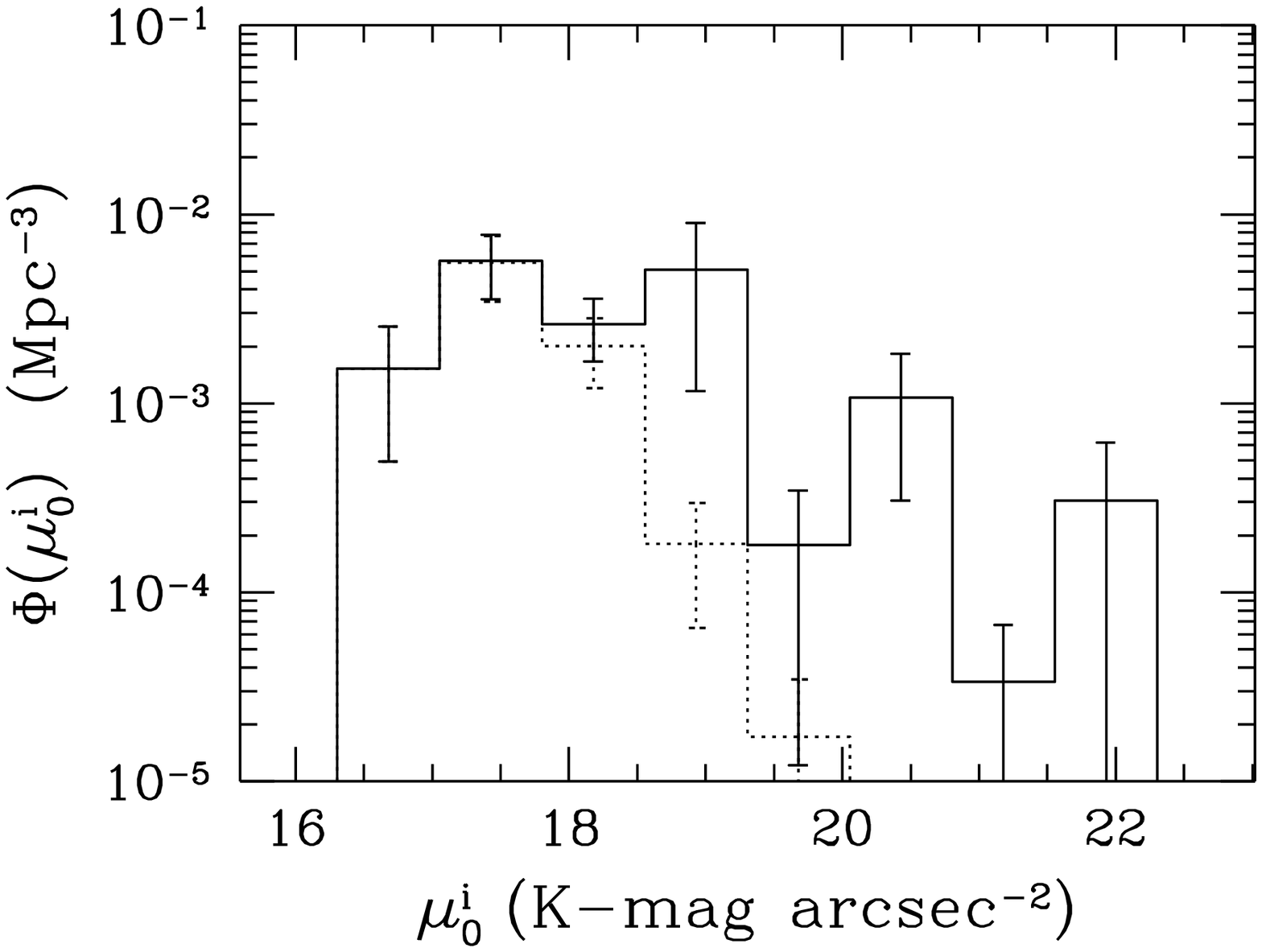}}
 \caption[]{
 The volume corrected distribution of the central surface brightness. 
The dashed line indicates the distribution for types earlier than type
T=6.  The number density is per bin size, which is in steps of 0.75
\magarc\ in $\mu_0$. 
 }
 \label{hiscs}
 \end{figure*}

\begin{figure*}
 \mbox{\epsfxsize=8.8cm\boundboxo{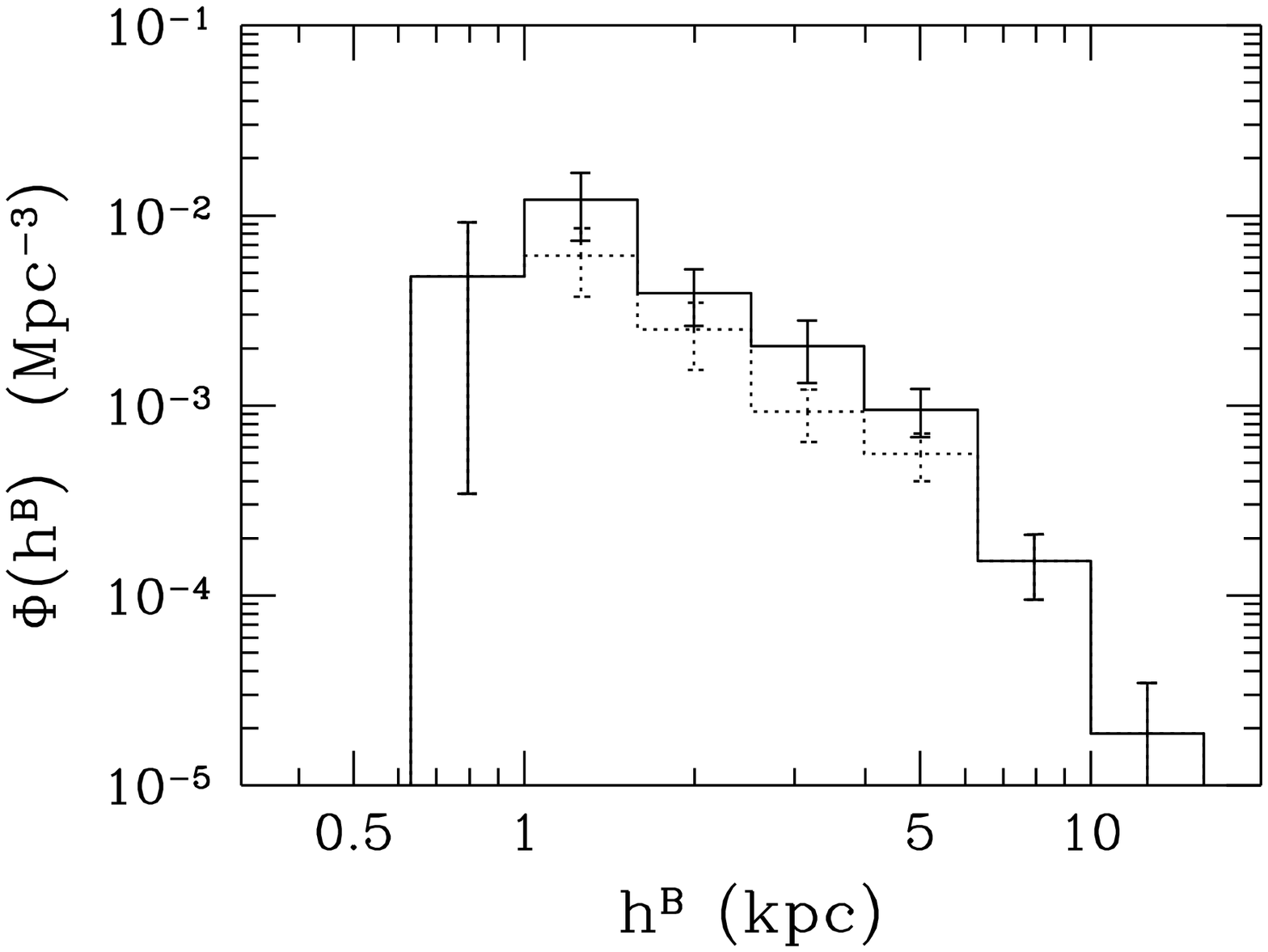}}
 \mbox{\epsfxsize=8.8cm\boundboxt{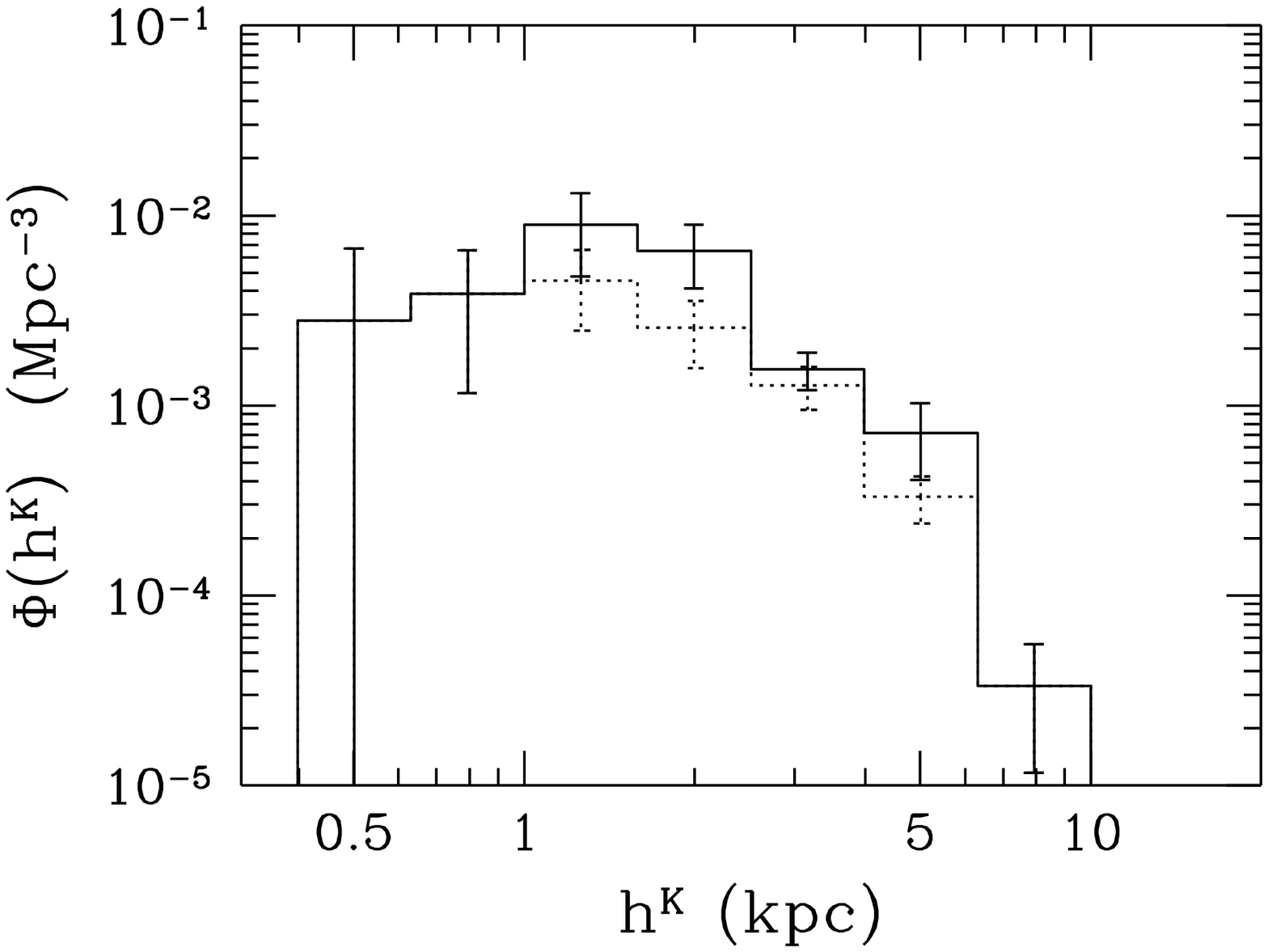}}
 \caption[]{
 The volume corrected distribution of the disk scalelengths.  The
dashed line indicates the distribution for type earlier than T=6.  The
number density is per bin size, which is in steps of 0.2 in log($h$). 
 }
 \label{hissc}
 \end{figure*}

}

% limit at 22.2 $K$-mag so relative volume
%(22.7-17)/(22.7-20)^3 ~12, we had ~6, so one in 2 finding one.  

By summing all bins in one direction, the bivariate distributions of
Fig.~\ref{bilogsccs} can be used to calculate the distributions of
\muo\ and $h$ separately. This figure indicates therefore where one can
expect problems in the determinations of the \muo\ and $h$
distributions due to the undersampling in the low surface brightness,
small scalelength region.  The \muo\ distributions will get incomplete
for central surface brightnesses fainter than 21.5 $B$-\magarc\ (19
$K$-\magarc) and the $h$ distributions should not be trusted for
scalelengths smaller than 1 kpc.  The undersampling in the \muo\
distribution is considerably reduced when only the galaxies with
scalelength larger than 1 kpc are used.  The undersampling problem of
this sample could also be circumvented by imposing absolute magnitude
or intrinsic diameter limits. To speak of {\em the} \muo\ distribution
is incorrect and one should indicate to what type of galaxies the
sample is restricted.

The distributions of central surface brightnesses of galaxies with
scalelength larger than 1 kpc are displayed in Fig.~\ref{hiscs}.  The
distributions are remarkably flat for the total sample.  The number
density density decreases by about a factor of 4 from \muo$^i \!\simeq
\!21$ to 24 $B$-\magarc\ and by a factor $\sim$10 from 17.5 to 22
$K$-\magarc. The distributions are narrower when only types earlier
than T=6 are used.  Disks of late-type galaxies are bluer, which makes
the overall distribution narrower in $B$ than in $K$.  The
distributions are not limited by selection effects at the bright end,
even if one assumes there is an upper limit to the total luminosity of
a galaxy (see Fig.~\ref{sclcs}). The number density of galaxies decreases
sharply with $\muo^i$ brighter than 20 $B$-\magarc\ ($\sim$16
$K$-\magarc).  At the faint end a limited volume is sampled, and
Fig.~\ref{sclcs} indicates that the sample is biased against galaxies
with a $\muo^i$ fainter than 23 $B$-\magarc\ even for galaxies with
scalelength larger than 1\,kpc.  Obviously galaxies with central surface
brightness fainter than 26 $B$-\magarc\ could never enter the sample. 
The distributions of \muo\ in Fig.~\ref{hiscs} could be slightly higher
at the faint end and should probably be extended to much lower
surface brightnesses. 

The volume corrected distributions of the logarithm of the scalelengths
(Fig.~\ref{hissc}) show first a small increase of galaxies to
scalelengths of about 1\,kpc.  This is probably caused by the
undersampling effect at low surface brightnesses and small
scalelengths. For scalelengths larger than 1\,kpc we notice a steady
decline of about a factor 100 in one dex. There is no segregation with
morphological type. 

\begin{figure*}
 \mbox{\epsfxsize=8.8cm\boundboxo{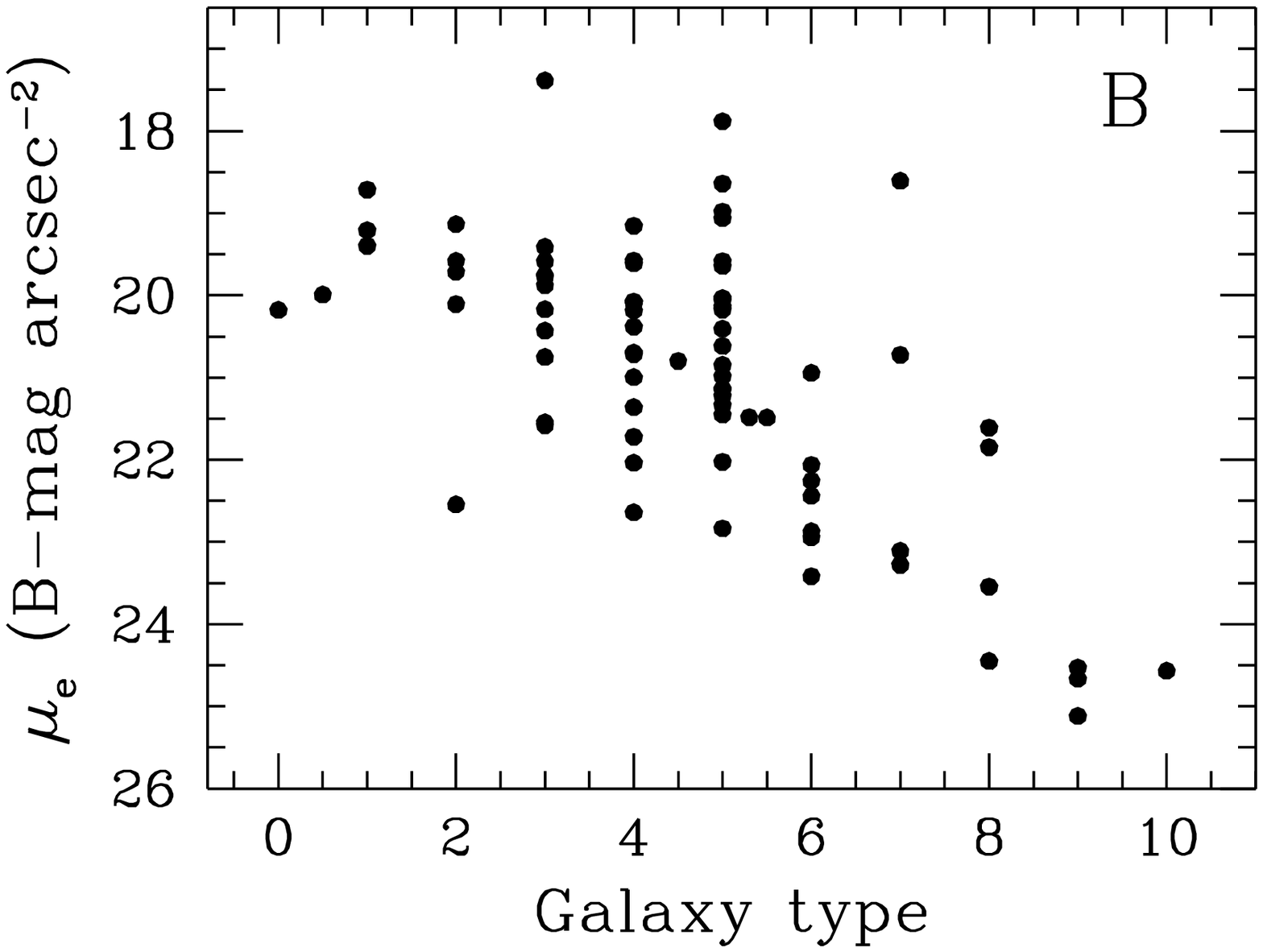}}
 \mbox{\epsfxsize=8.8cm\boundboxt{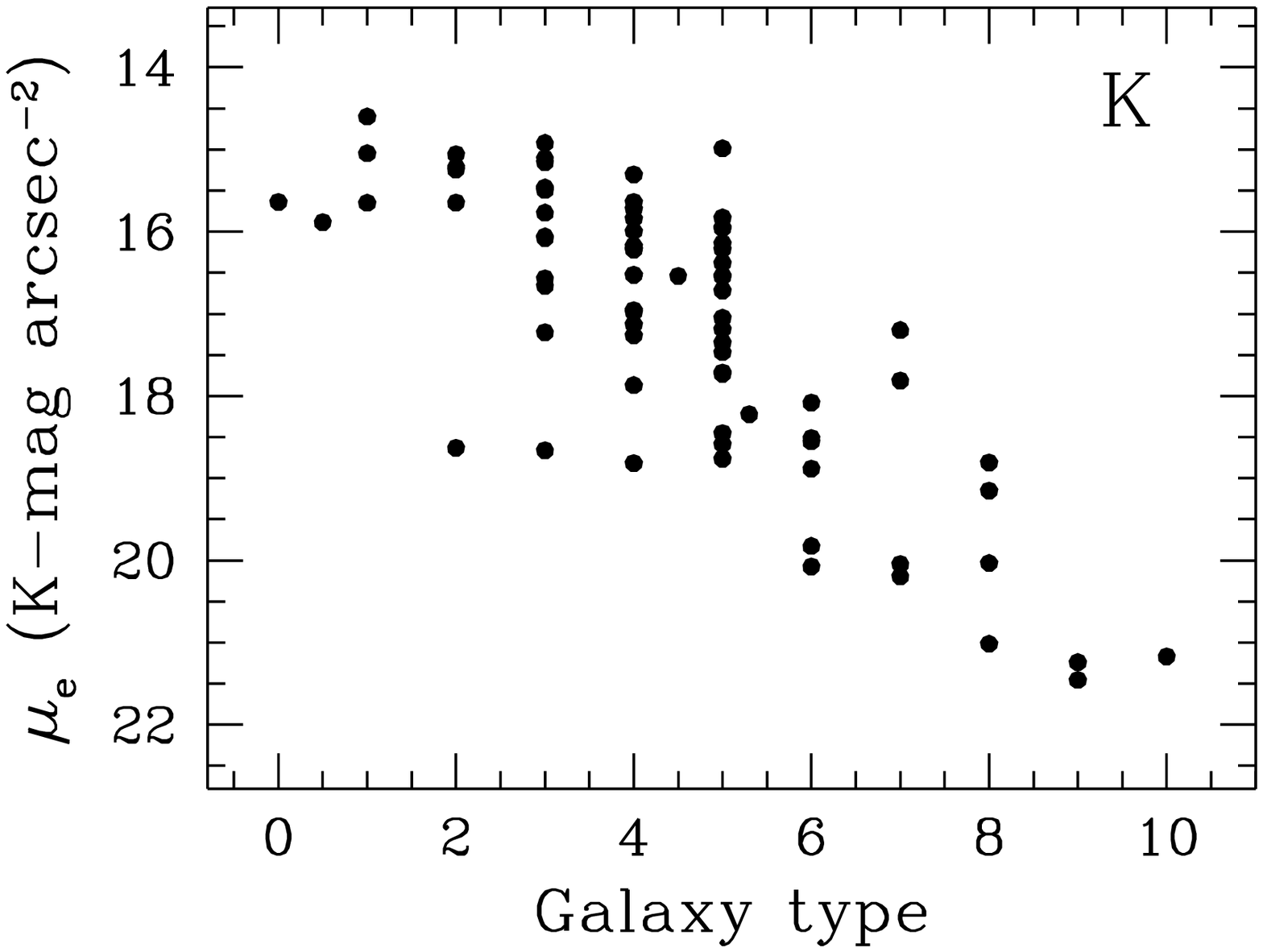}}
 \caption[]{
The Galactic extinction corrected effective surface brightness of the
bulge as function of morphological RC3 type. 
}
 \label{type_es}
 \end{figure*}

\begin{figure*}
 \mbox{\epsfxsize=8.8cm\boundboxo{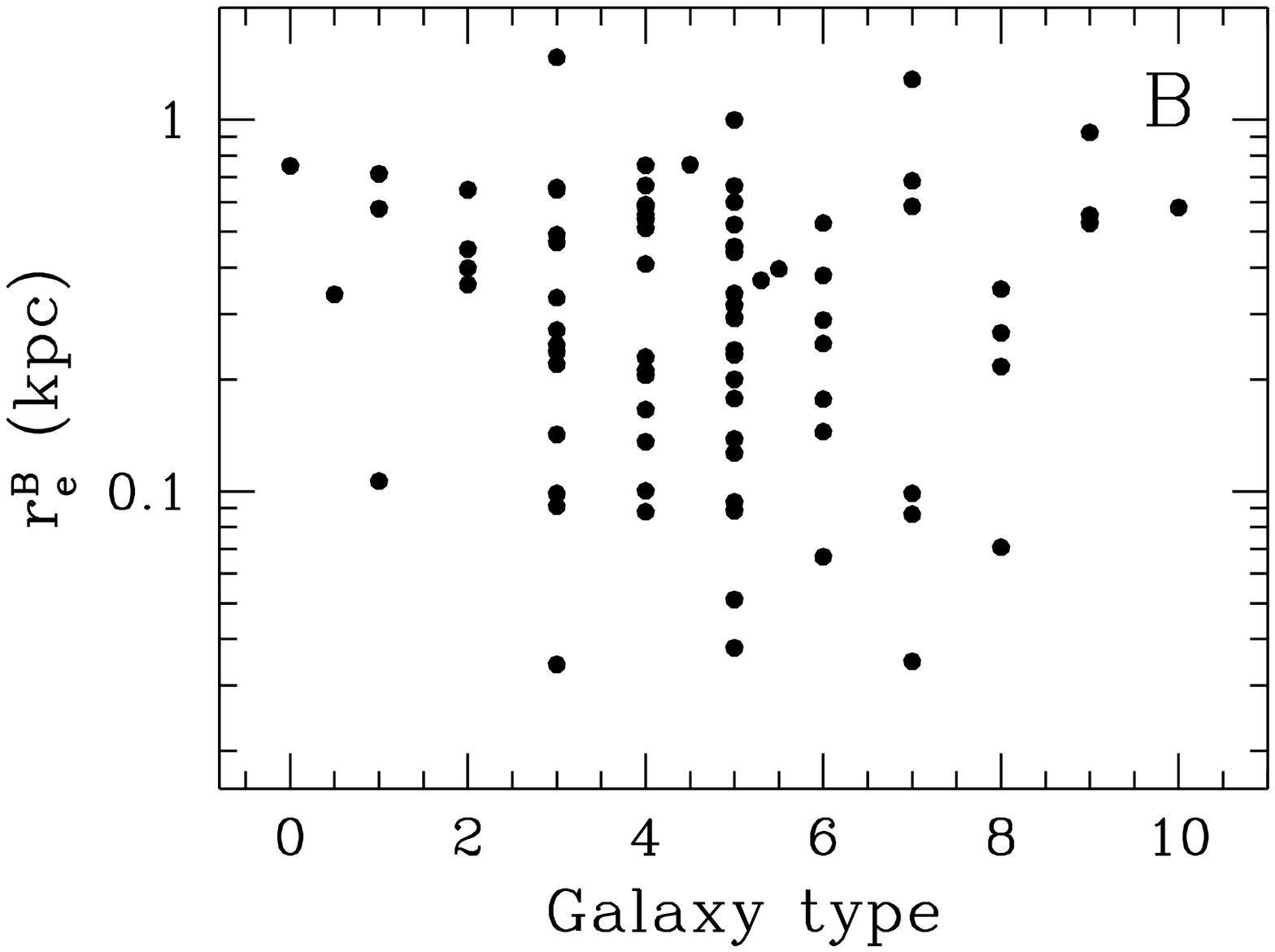}}
 \mbox{\epsfxsize=8.8cm\boundboxt{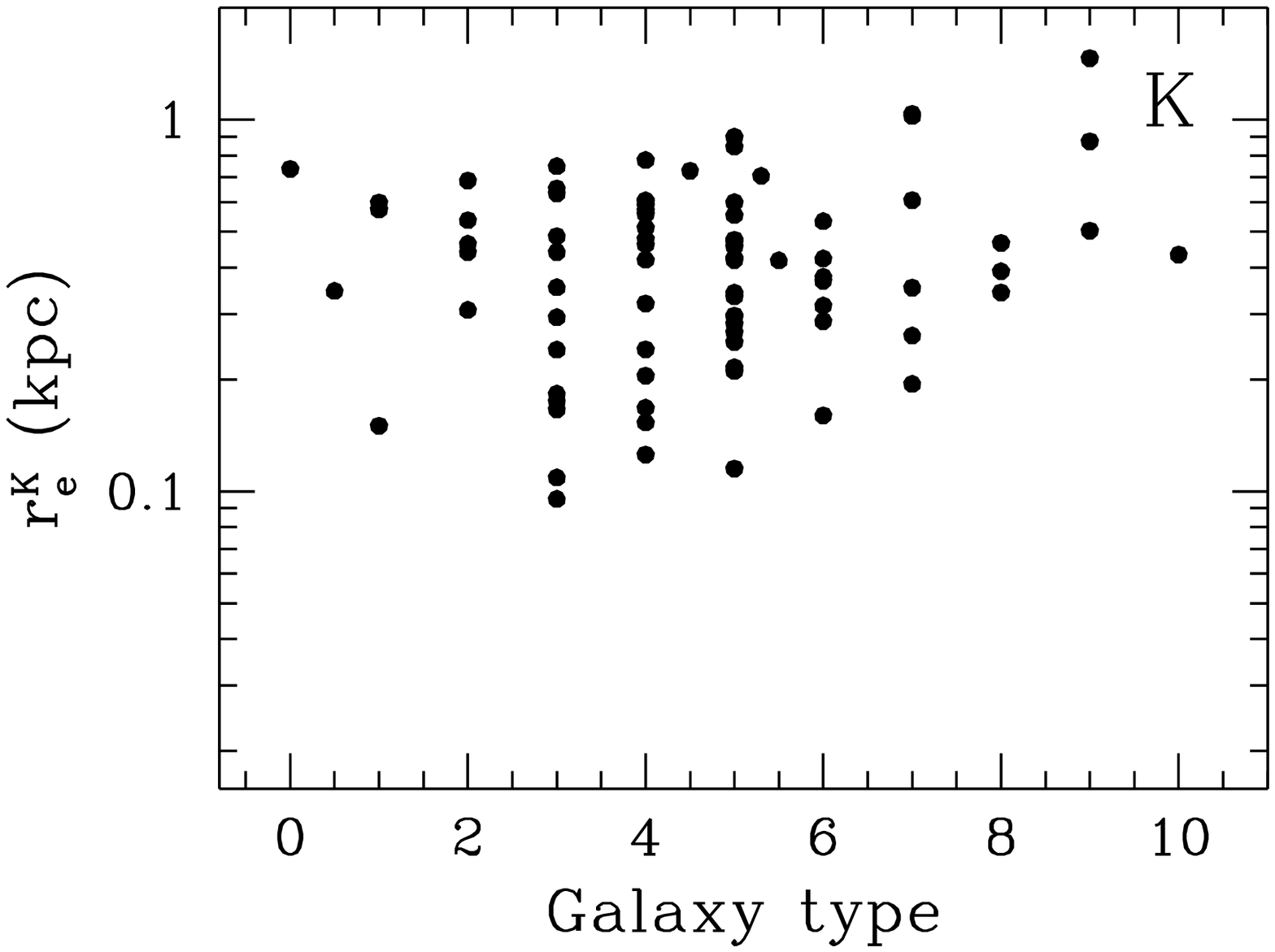}}
 \caption[]{
The effective radius of the bulge as function of morphological type.
}
 \label{type_er}
 \end{figure*}

The most important results obtained in this subsection are as follows.
There is a large range in disk central surface brightnesses among
galaxies, mainly due to the lower surface brightnesses of late-type
galaxies. The range decreases slightly when a transparent inclination
correction is used. Selection effects are very significant in
determining number density distributions and after correcting for these
effects there is no single preferred value for the central surface
brightnesses of disks. There is an upper limit to \muo, but the number
density distribution decreases only slowly at the faint end.

\subsection{The bulge parameters}

The same diagrams used to describe the disk parameters are now used to
present the bulge parameters.  The distributions of effective surface
brightness are presented in Fig.~\ref{type_es}.  The effective surface
brightness shows a tight correlation with type index, especially
considering the uncertainty of at least 1.5 T-units (1 sigma) in type
index (Lahav~\cite{Lah95}).  Almost all of the scatter can be explained
by this uncertainty.  This relation also holds for the \mue\ parameters
obtained with the other fitting methods presented in Paper~II, although
with a slightly larger scatter.  There is no apparent correlation of
effective radius with galaxy type (Fig.~\ref{type_er}).  The relations
in Figs~\ref{type_es} and \ref{type_er} are tighter in $K$ than in $B$. 
There are several explanations for this effect.  Bulges are relatively
brighter with respect to the disks in $K$ compared to $B$, which will
make the fit routine work better.  Furthermore, circumnuclear star
formation and dust lanes will affect the $B$ passband more than the $K$
passband and make the quality of the decomposition worse.  There are
some galaxies in the sample with clear circumnuclear star formation and
with dust lanes right down to the center.  Finally, there is the effect
of the Freeman Type~II profiles (Freeman~\cite{Freeman}) which is
reduced in $K$, thus making fitting easier (see Paper~II).  The
distribution of points in the (\mue,\re)-plane (Fig.~\ref{eres}) shows
no correlation.  The absence of a correlation between \re\ and
morphological type makes the trend in the distribution of the absolute
bulge magnitude ($M_{\rm bulge} \!\propto \!\mue \!- \!2.5\log(\re^2)$)
versus type (Fig.~\ref{type_Mb}) dominated by the \mue. 
 %The scatter is rather large though. 

{
\begin{figure*}
 \mbox{\epsfxsize=8.8cm\boundboxo{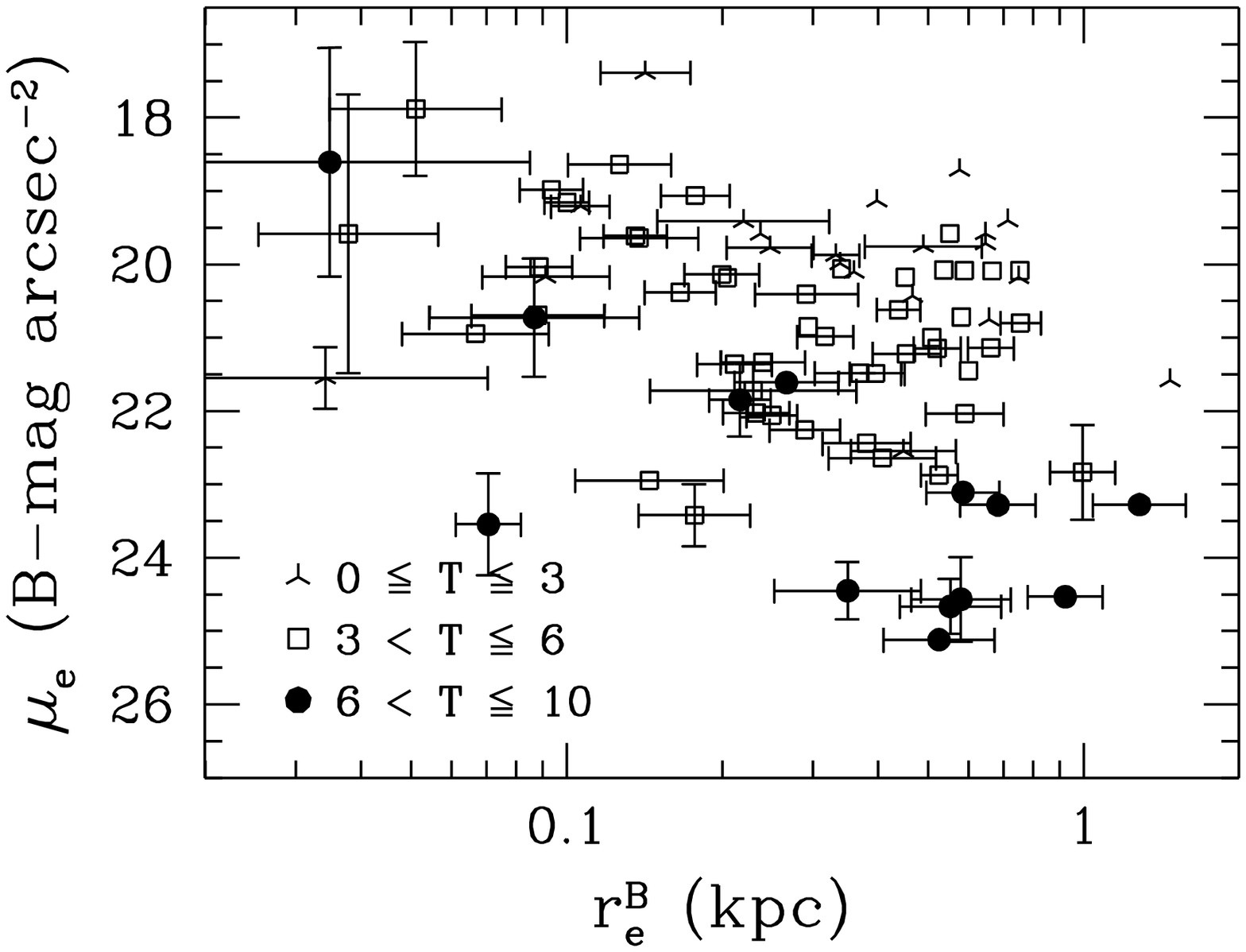}}
 \mbox{\epsfxsize=8.8cm\boundboxt{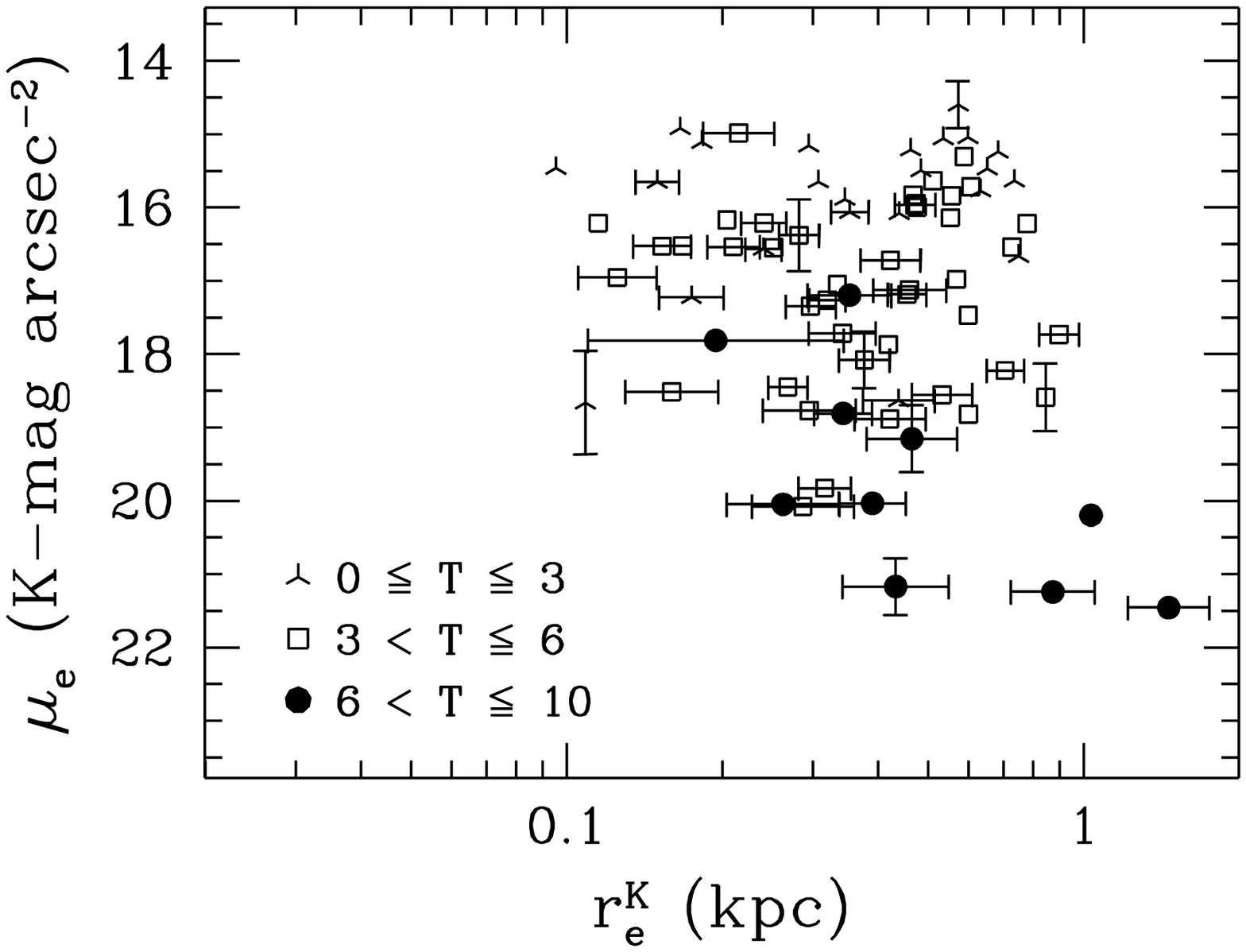}}
 \caption[]{
The effective radius of the bulge versus the effective surface brightness 
at this radius. Different symbols are used to denote the indicated type 
ranges.
}
 \label{eres}
 \end{figure*}

\begin{figure*}
 \mbox{\epsfxsize=8.8cm\boundboxo{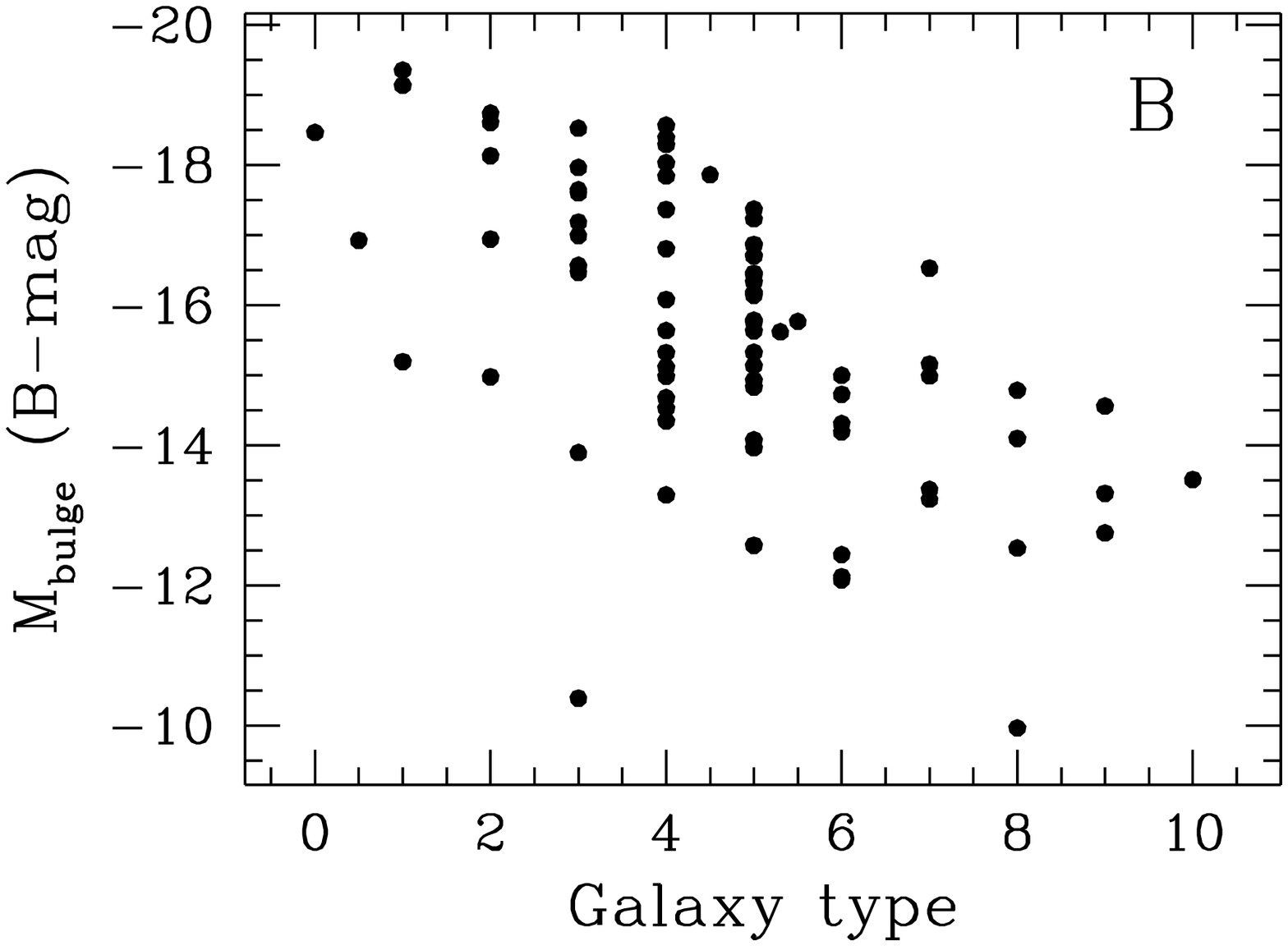}}
 \mbox{\epsfxsize=8.8cm\boundboxt{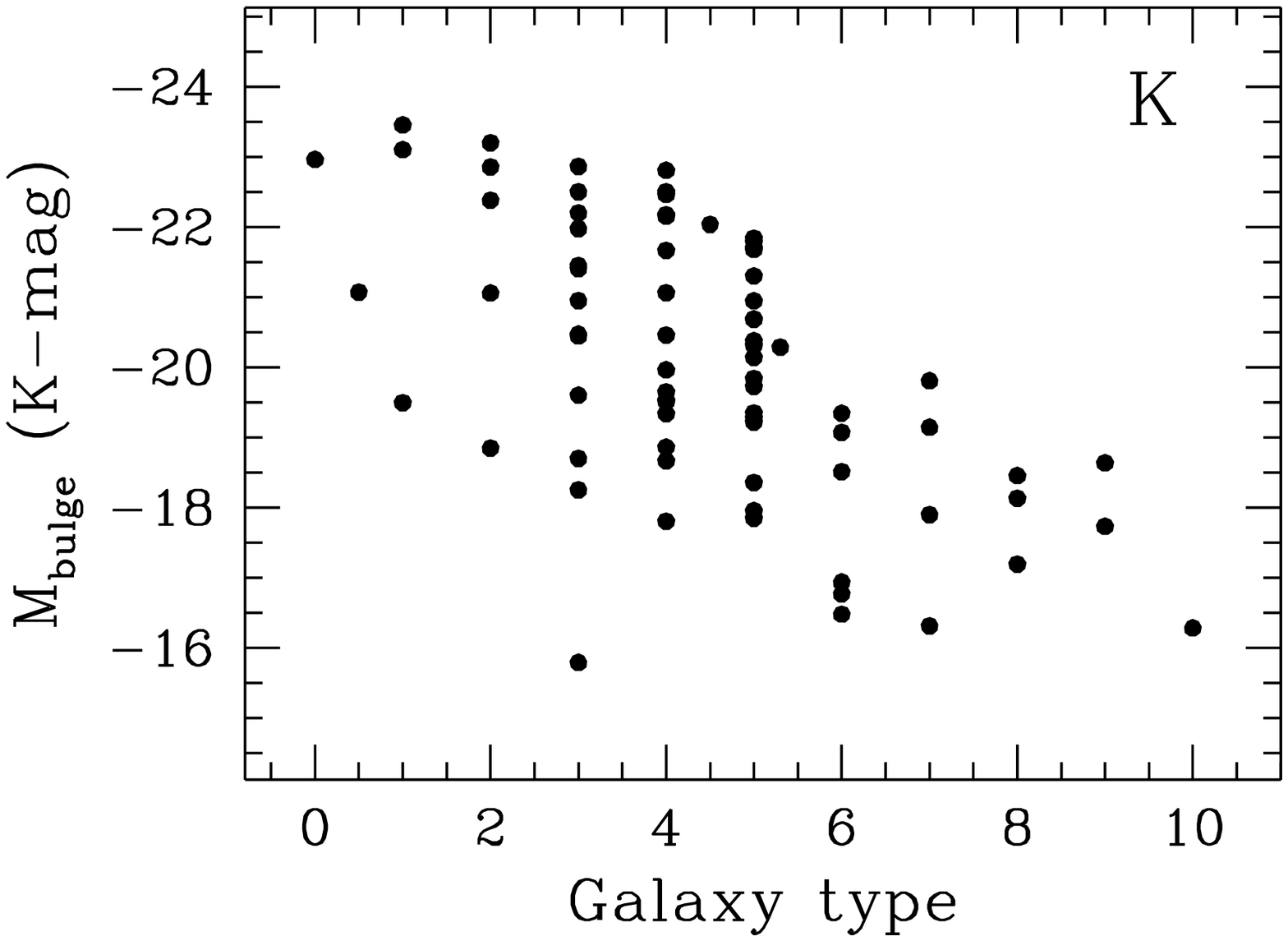}}
 \caption[]{
The Galactic extinction corrected absolute magnitude of the
bulge as function of morphological RC3 type. 
}
 \label{type_Mb}
 \end{figure*}

\begin{figure*}
 \mbox{\epsfxsize=8.8cm\epsfbox[87 140 482 440]{pbfbul.cps}}
 \mbox{\epsfxsize=8.8cm\epsfbox[67 140 462 440]{pkfbul.cps}}
 \caption[]{
 The volume corrected bivariate distribution of galaxies in the
($\mu_{\rm e}$,$r_{\rm e}$)-plane.  The number density $\Phi(\mue$,\re) is per bin size,
which is in steps of 0.3 in log($r_{\rm e}$) and 2 \magarc\ in $\mu_{\rm
e}$. 
 }
 \label{bilogeres}
 \end{figure*}

\begin{figure*}
 \mbox{\epsfxsize=8.8cm\boundboxo{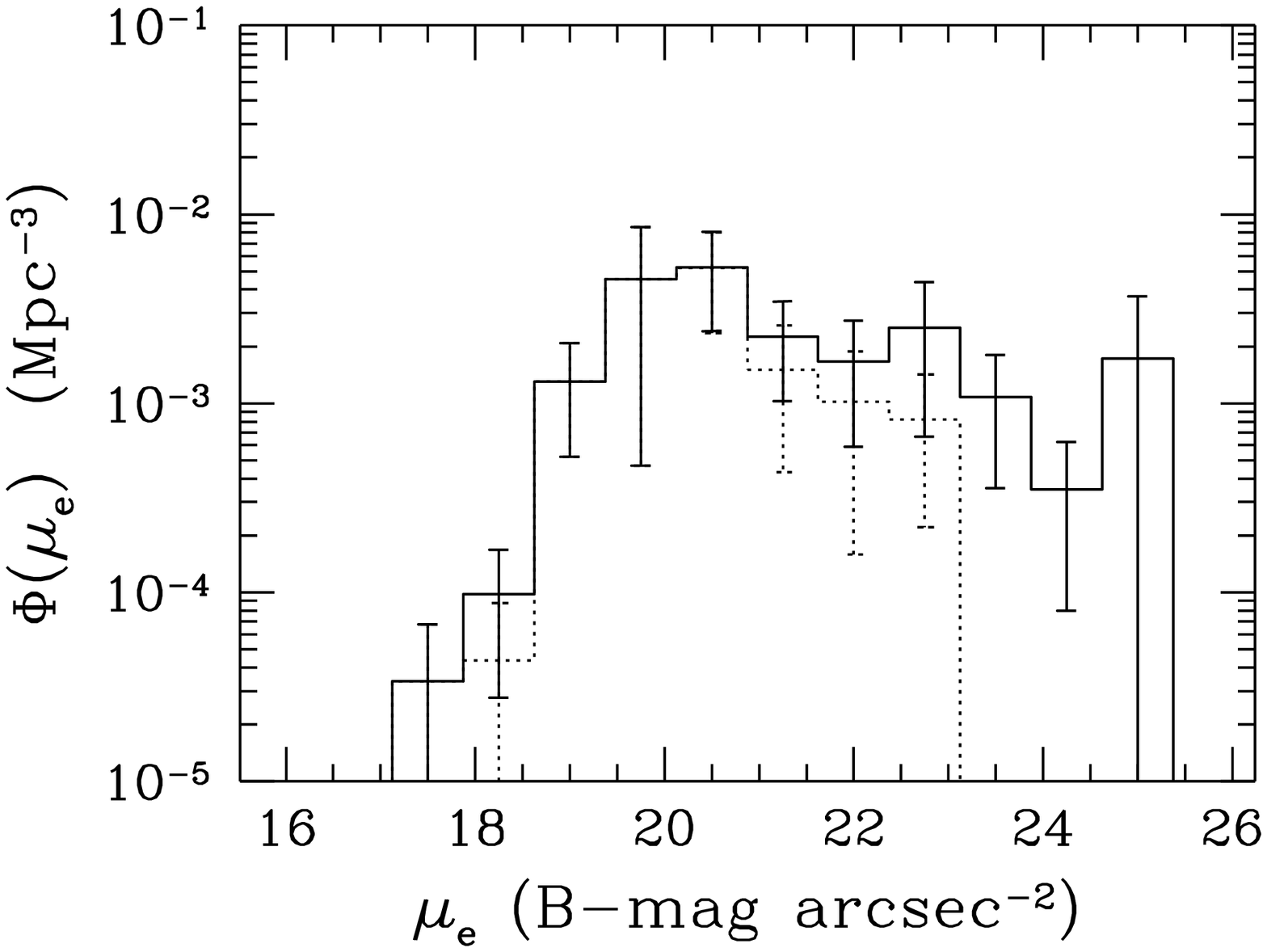}}
 \mbox{\epsfxsize=8.8cm\boundboxt{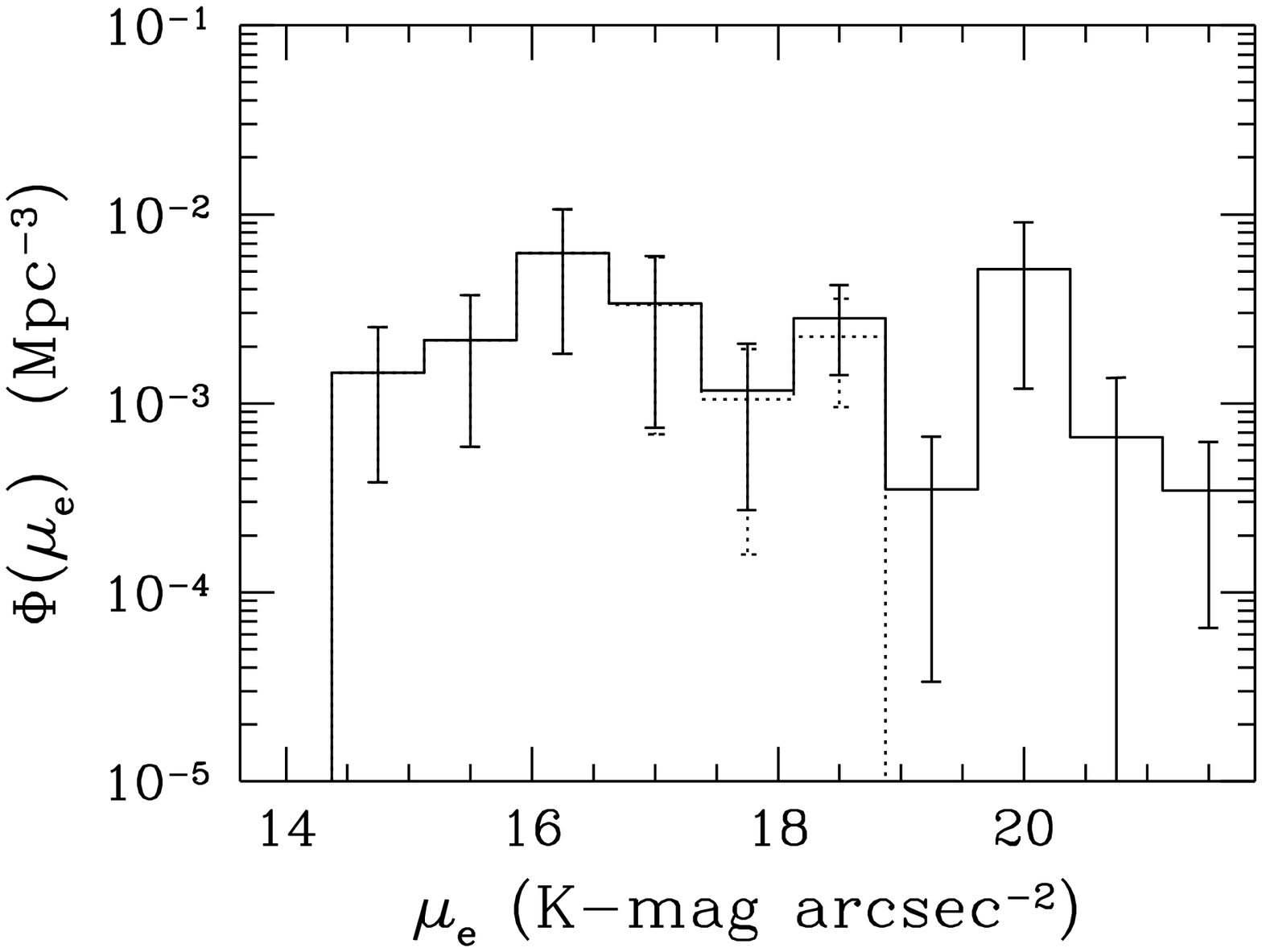}}
 \caption[]{
 The volume corrected distribution of the effective surface brightnesses
of the bulge.  The dashed line indicates the distribution for types
with T$<$6.  The $\Phi$ distribution is per bin size, which is in
steps of 0.75 \magarc\ in $\mu_{\rm e}$. 
 }
 \label{hises}
 \end{figure*}

\begin{figure*}
 \mbox{\epsfxsize=8.8cm\boundboxo{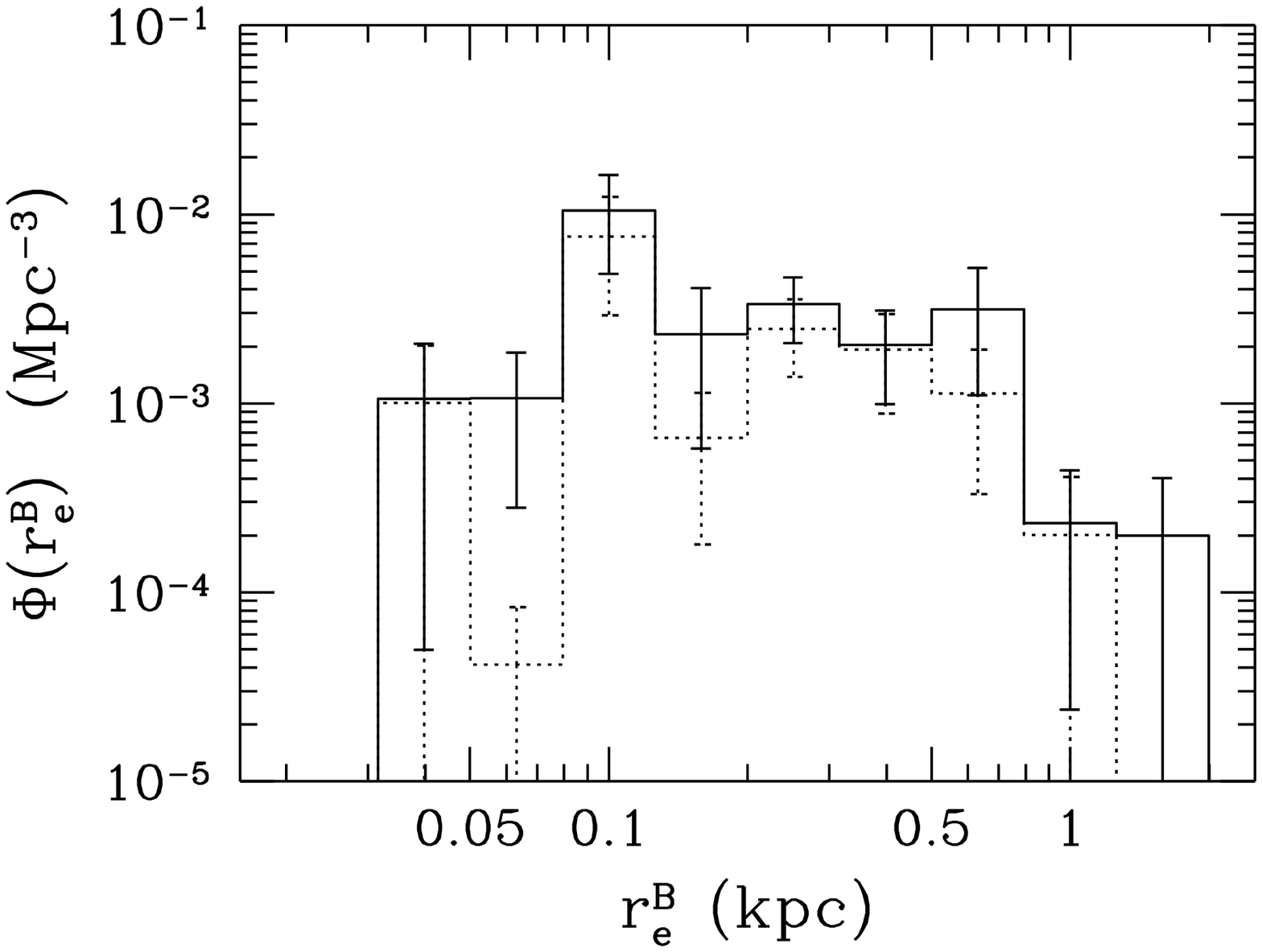}}
 \mbox{\epsfxsize=8.8cm\boundboxt{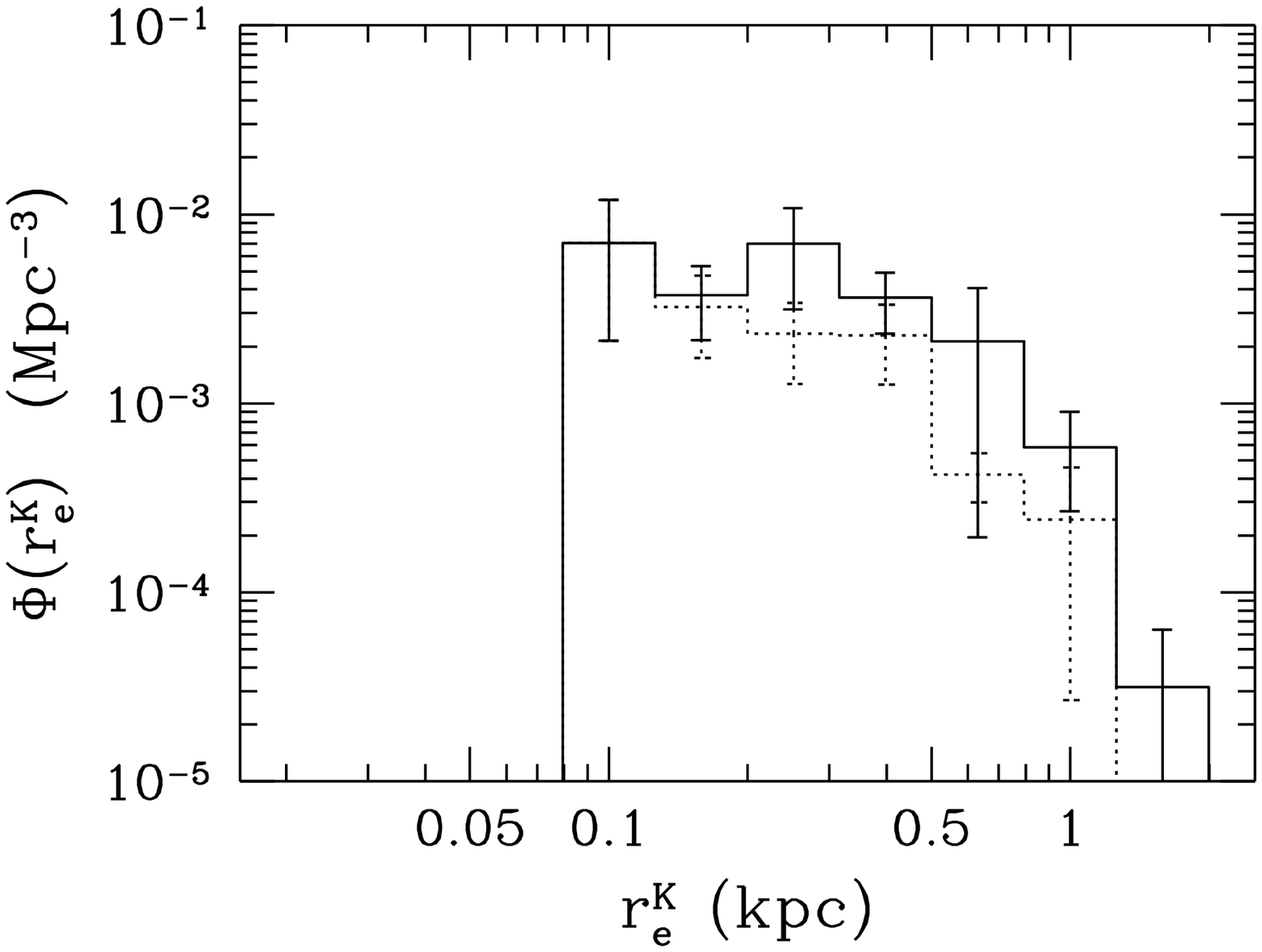}}
 \caption[]{
 The volume corrected distribution of the bulge effective radii.  The
dashed line indicates the distribution for types with T$<$6.  The
$\Phi$ distribution is per bin size, which is in steps of 0.2 in
log($r_{\rm e}$). 
 }
 \label{hiser}
 \end{figure*}

}

The bivariate distribution of \mue\ and \re\ (Fig.~\ref{bilogeres})
shows no trends.  The dominant type of galaxy in our local universe has
a bulge with effective radius in the range of 0.1-0.3~kpc and effective
surface brightness of order 21 $B$-\magarc\ ($\sim$16 $K$-\magarc). 
The relation between the bulge parameters and the diameter selection
criterion is not very obvious and therefore all galaxies are used in
the calculations of the separate bulge parameter distributions. The
volume corrected distributions of the bulge \mue\ and \re\
(Figs~\ref{hises} and \ref{hiser}) show the same behavior as the disk
parameters, i.e.\ constant distribution of the effective surface
brightness and a steady decline of a factor of 50-100 per dex of the
effective radius in the $K$ passband.

%\subsection{The bar parameters}
%
%
%  \input{type_bc}
%  \input{type_ma}
%  \input{type_mima}
%
%The bar parameters are uncertain and they should not be investigated
%without keeping the relation to the disk and bulge parameters into
%account.  Both the central surface brightness of the bar
%(Fig.~\ref{type_bc}) and bar major axis length (Fig.~\ref{type_ma})
%show little correlation with morphological type.  The axis ratio of the
%bar does show a weak trend with morphological type
%(Fig.~\ref{type_mima}) in the sense that later type galaxies have on
%average a more elongated bar.

\subsection{The bulge/disk relation}

%The relation between bulge and disk parameters is important. 
%Even though they might originate from different epochs of galaxy
%formation, they are related to each other, 
 %(Eggen et al.~\cite{Egg62}; refs!) 
% because they sit in the same potential well.
 %, which will influence them both. 
 The chronology of the bulge and disk formation is a major issue and
the relationships between bulge and disk parameters might give some
insight in this matter.  A strong correlation between bulge and disk
parameters is expected if the bulge formed from the disk by secular
evolution.  A correlation might be expected in the hierarchical infall
and small merger models producing bulges, because both bulge and disk
originate from the same smaller components. In models where the bulge
forms first and the  disk forms later, there is no obvious reason for a
bulge-disk correlation. 

\begin{figure*}
 \mbox{\epsfxsize=8.8cm\boundboxo{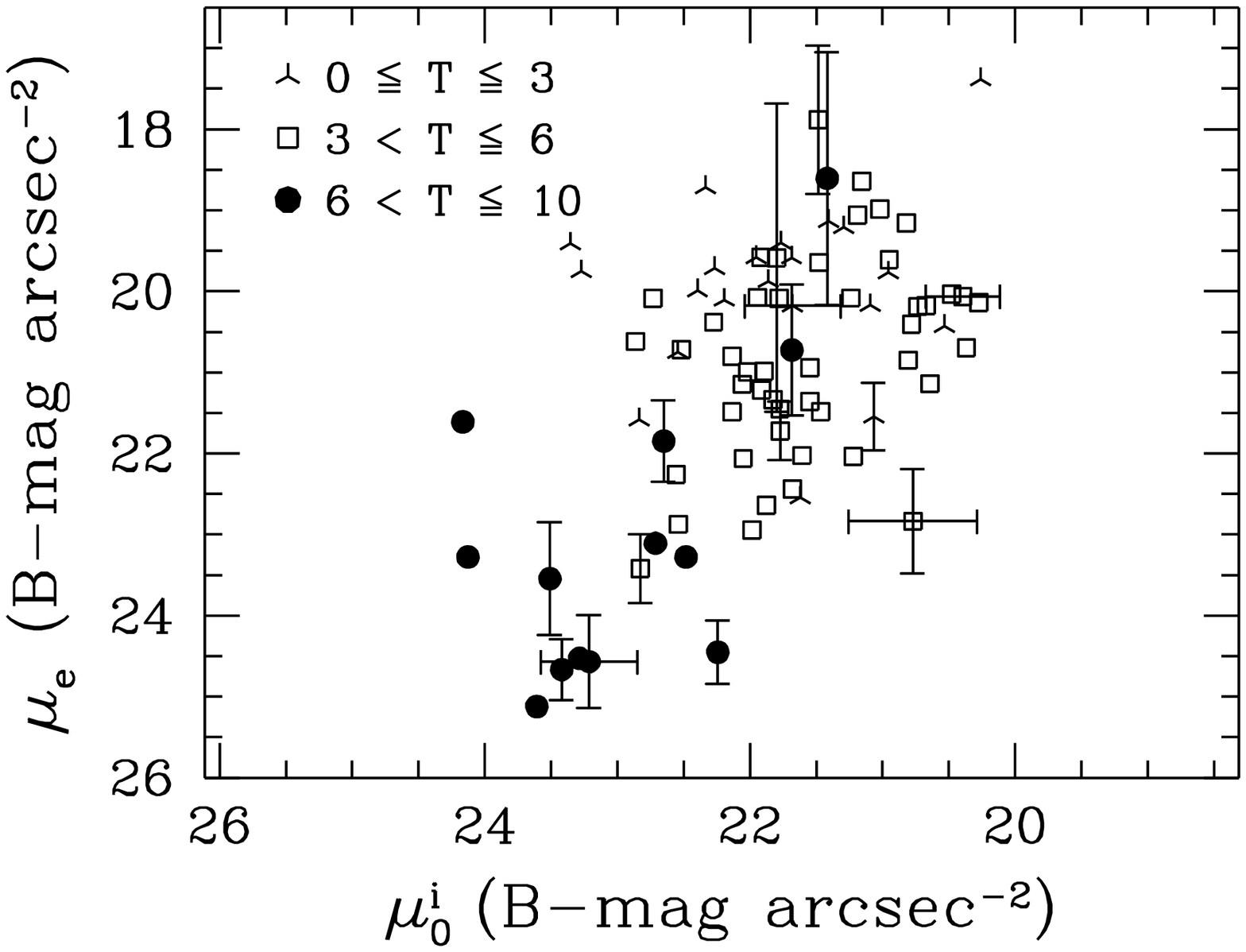}}
 \mbox{\epsfxsize=8.8cm\boundboxt{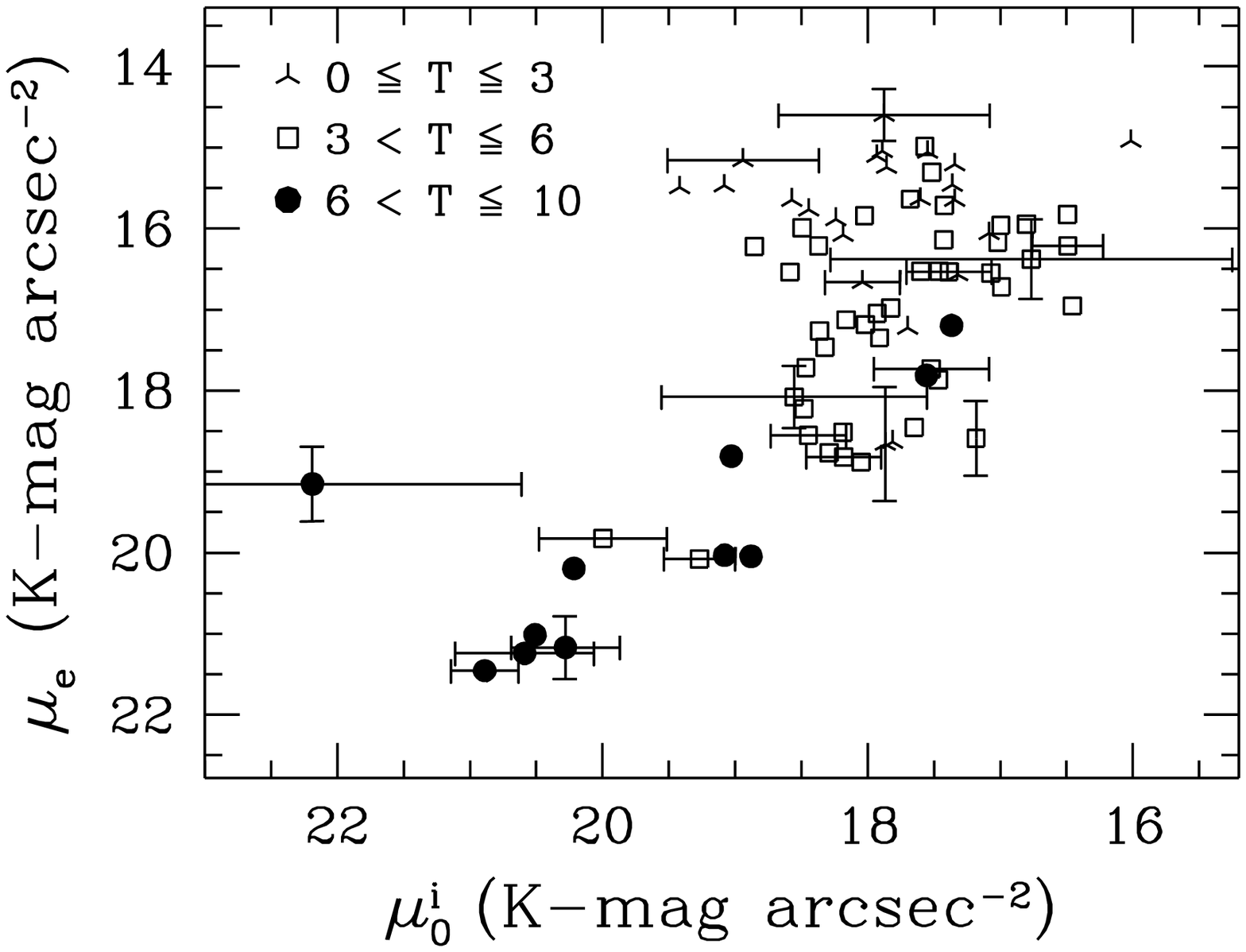}}
 \caption[]{
The central surface brightness of the disk versus the effective surface
brightness of the bulge. 
Different symbols are used to denote the indicated morphological type
ranges. 
}
 \label{cses}
 \end{figure*}

\begin{figure*}
 \mbox{\epsfxsize=8.8cm\boundboxo{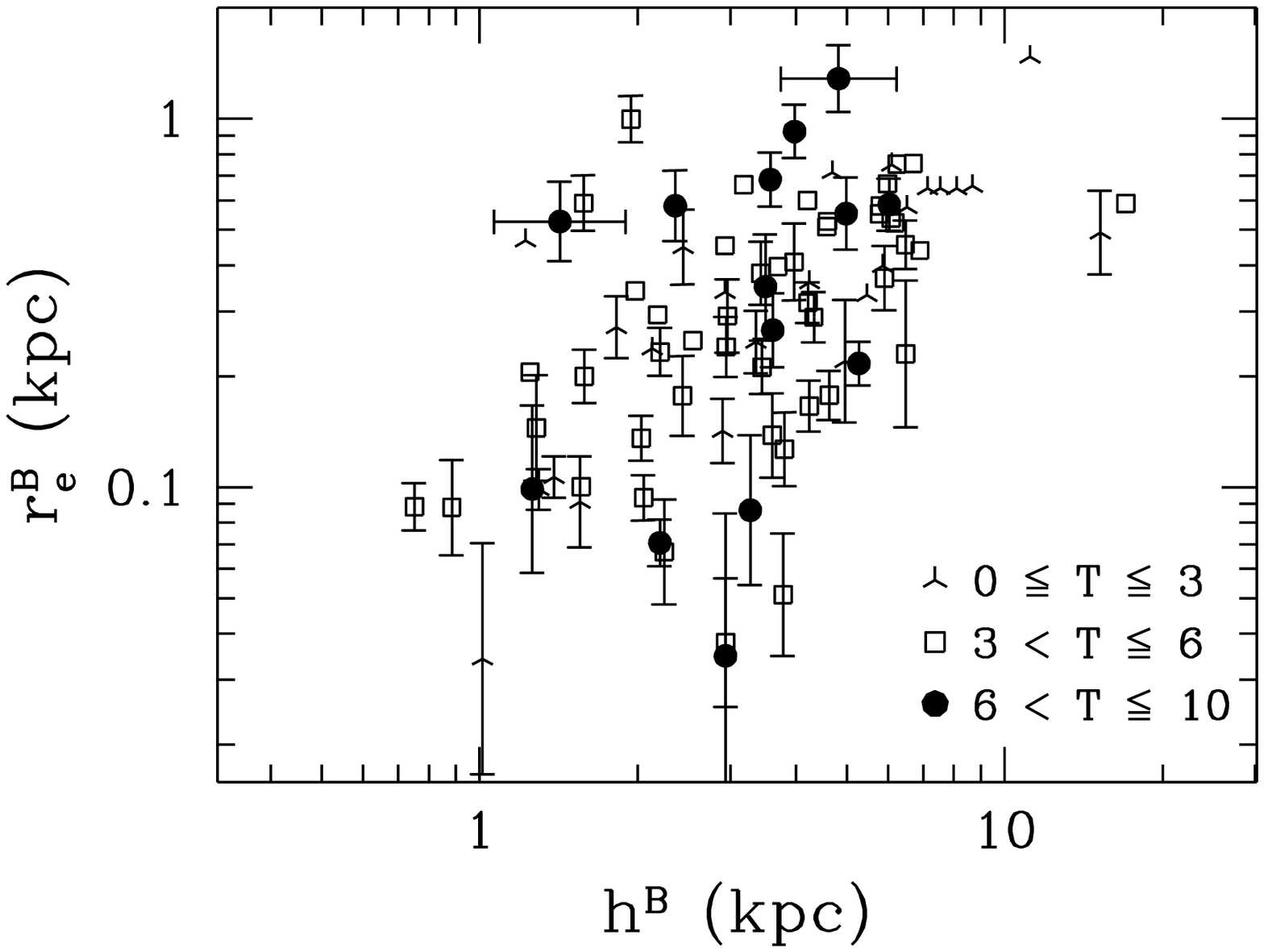}}
 \mbox{\epsfxsize=8.8cm\boundboxt{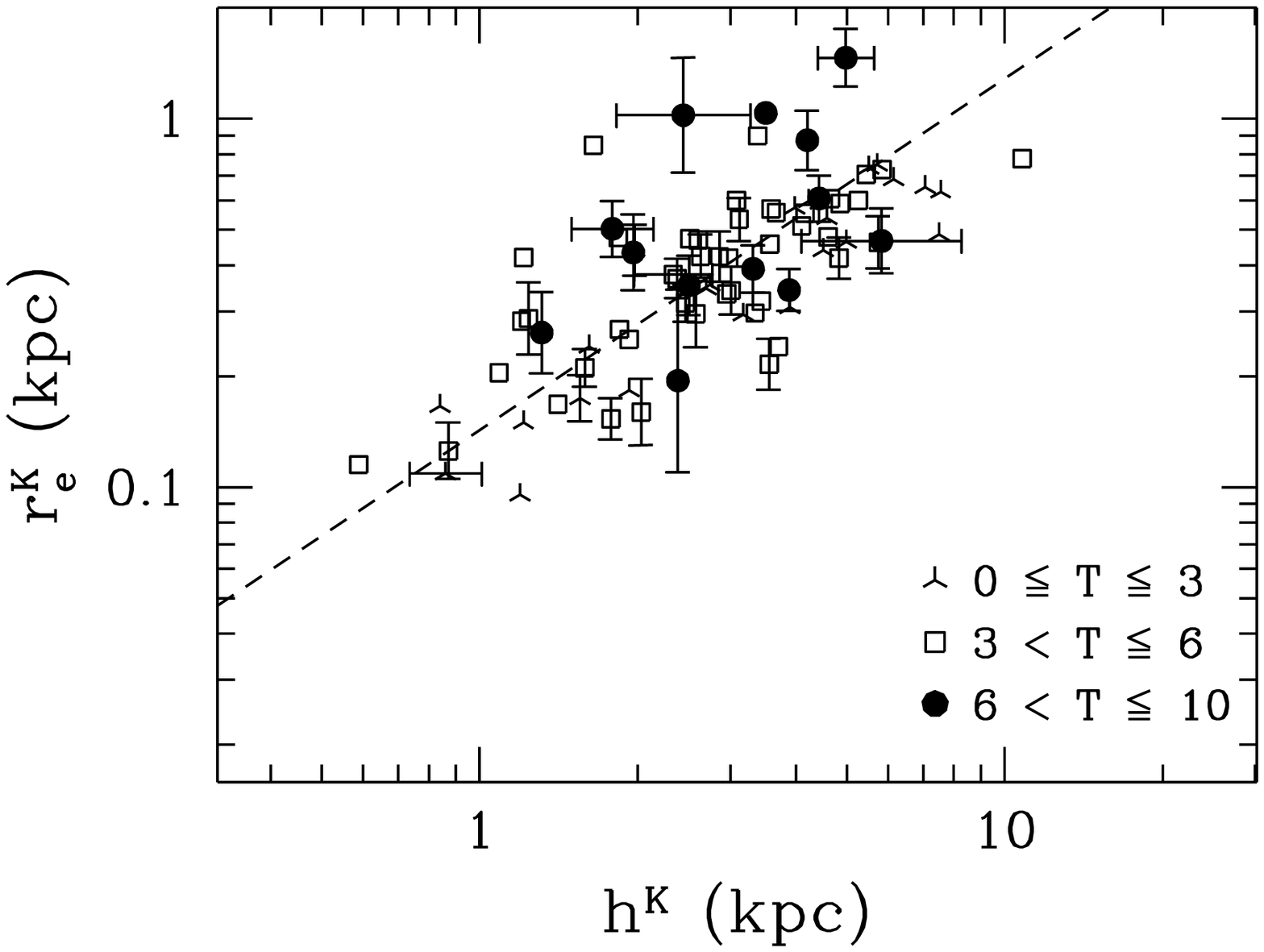}}
 \caption[]{
 The scalelength of the disks versus the effective radius of the bulge. 
Different symbols are used to denote the indicated morphological type
ranges.  The dashed line in the $K$ passband diagram gives the least
squares fit relationship between both parameters. 
 }
 \label{scer}
 \end{figure*}

\begin{figure*}
 \mbox{\epsfxsize=8.8cm\boundboxo{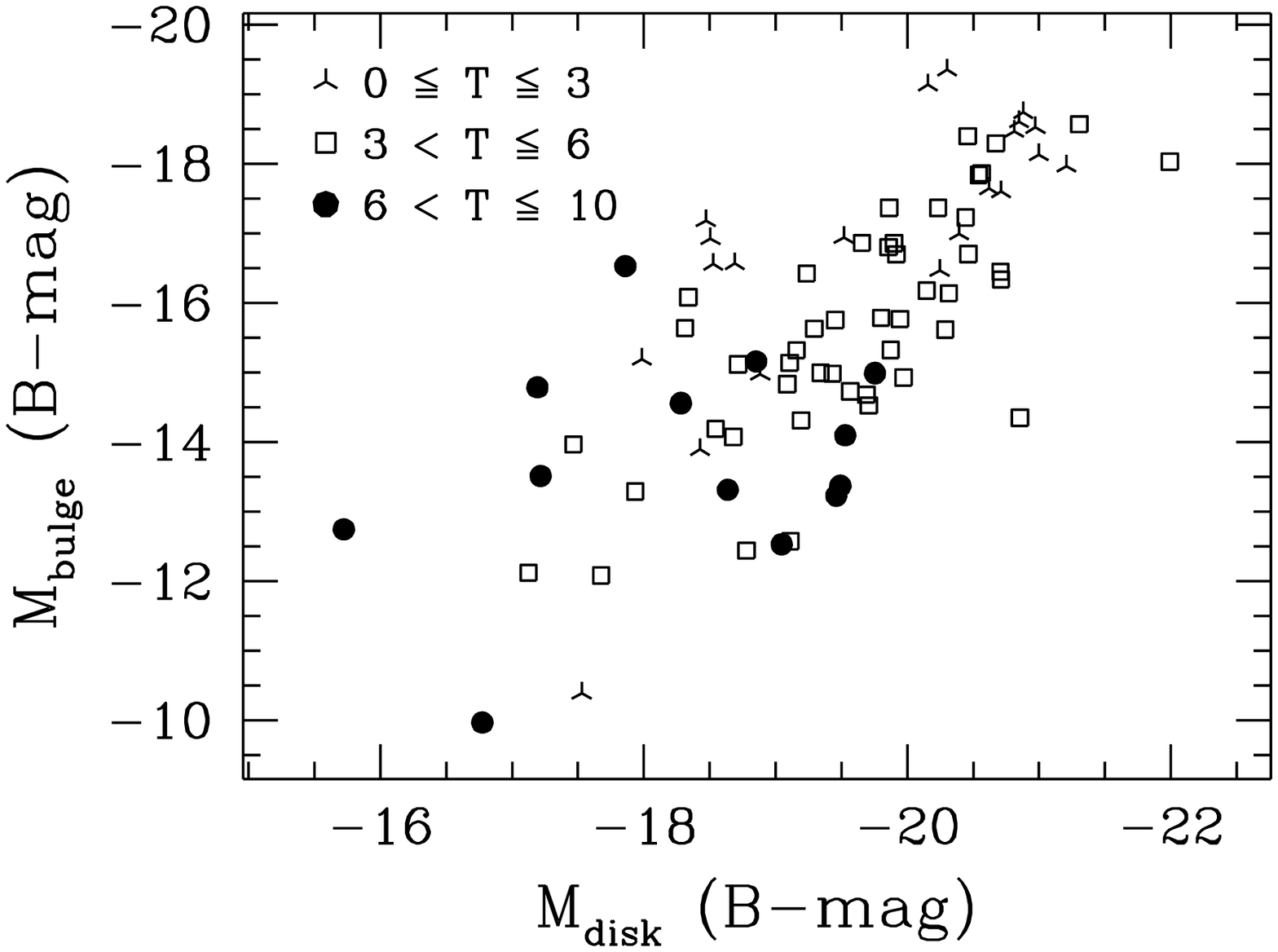}}
 \mbox{\epsfxsize=8.8cm\boundboxt{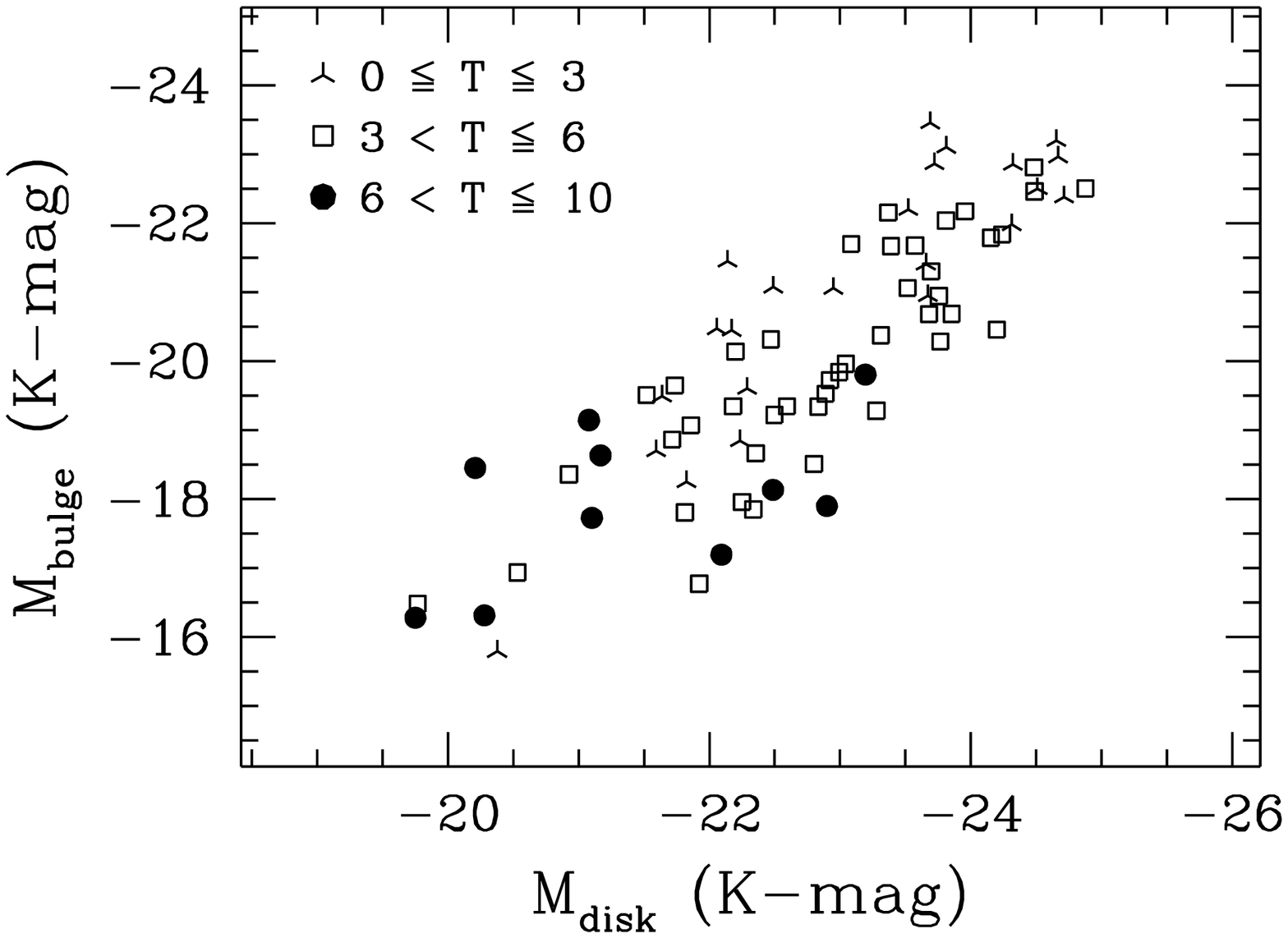}}
 \caption[]{
The galactic absorption corrected absolute magnitude of the
disk versus that of bulge. Morphological types ranges are denoted by the
indicated symbols.
}
 \label{MdMb}
 \end{figure*}

\begin{figure*}
 \mbox{\epsfxsize=8.8cm\boundboxo{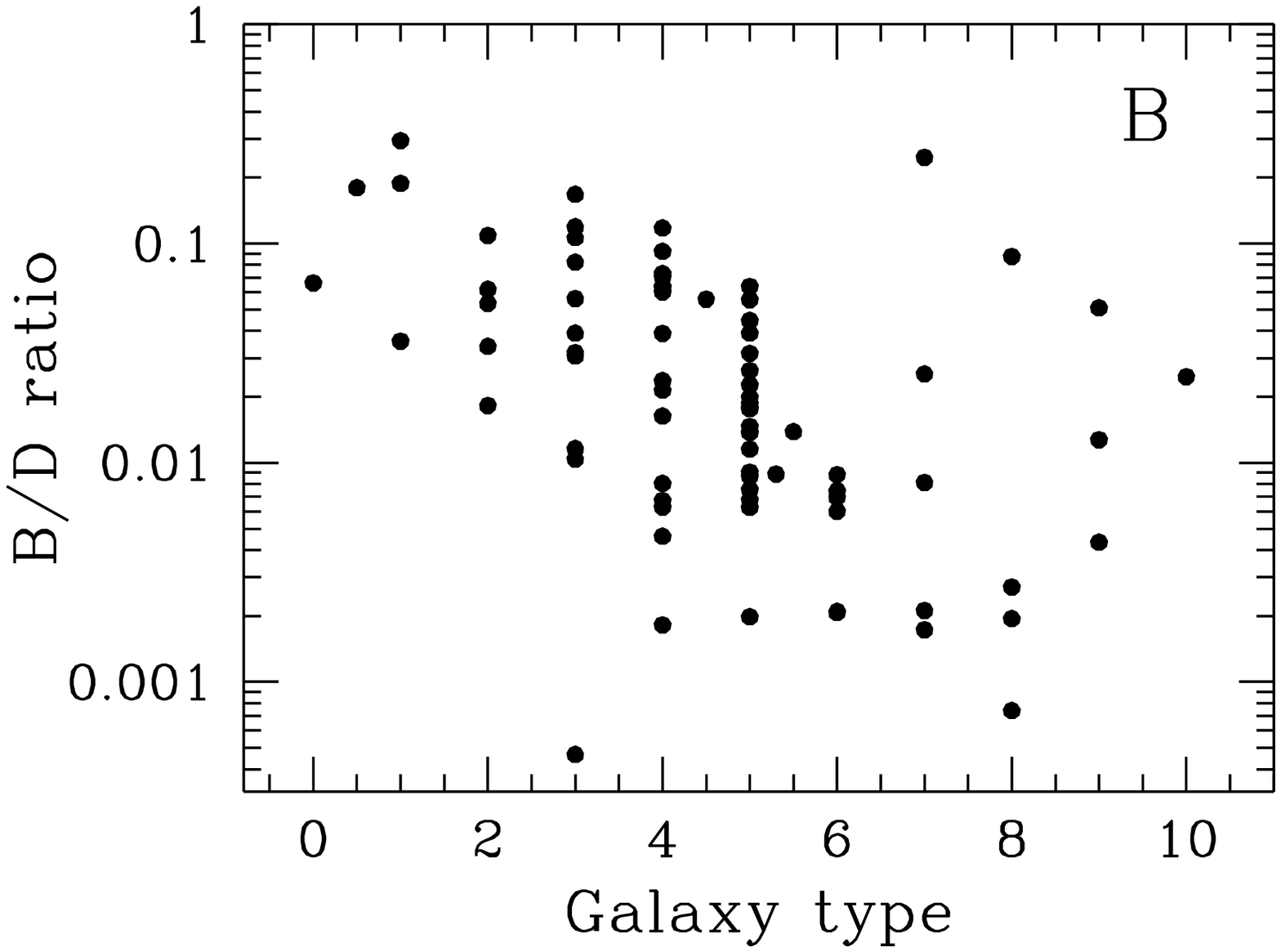}}
 \mbox{\epsfxsize=8.8cm\boundboxt{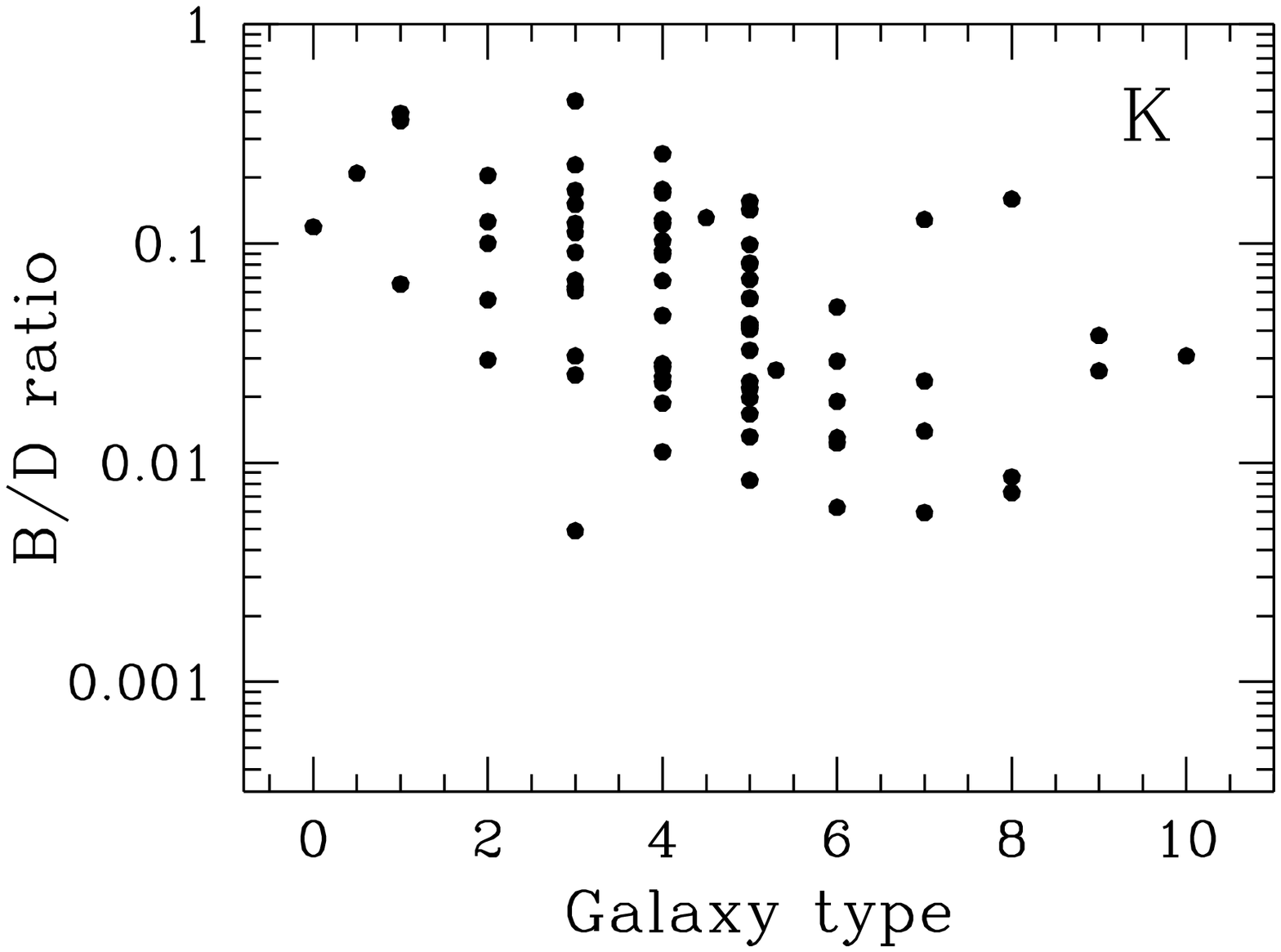}}
 \caption[]{
The bulge to disk ratio as function of morphological type, using the results
from the 2D fit of Paper~II.
 }
 \label{type_lBD}
 \end{figure*}

\begin{figure*}
 \mbox{\epsfxsize=8.8cm\boundboxo{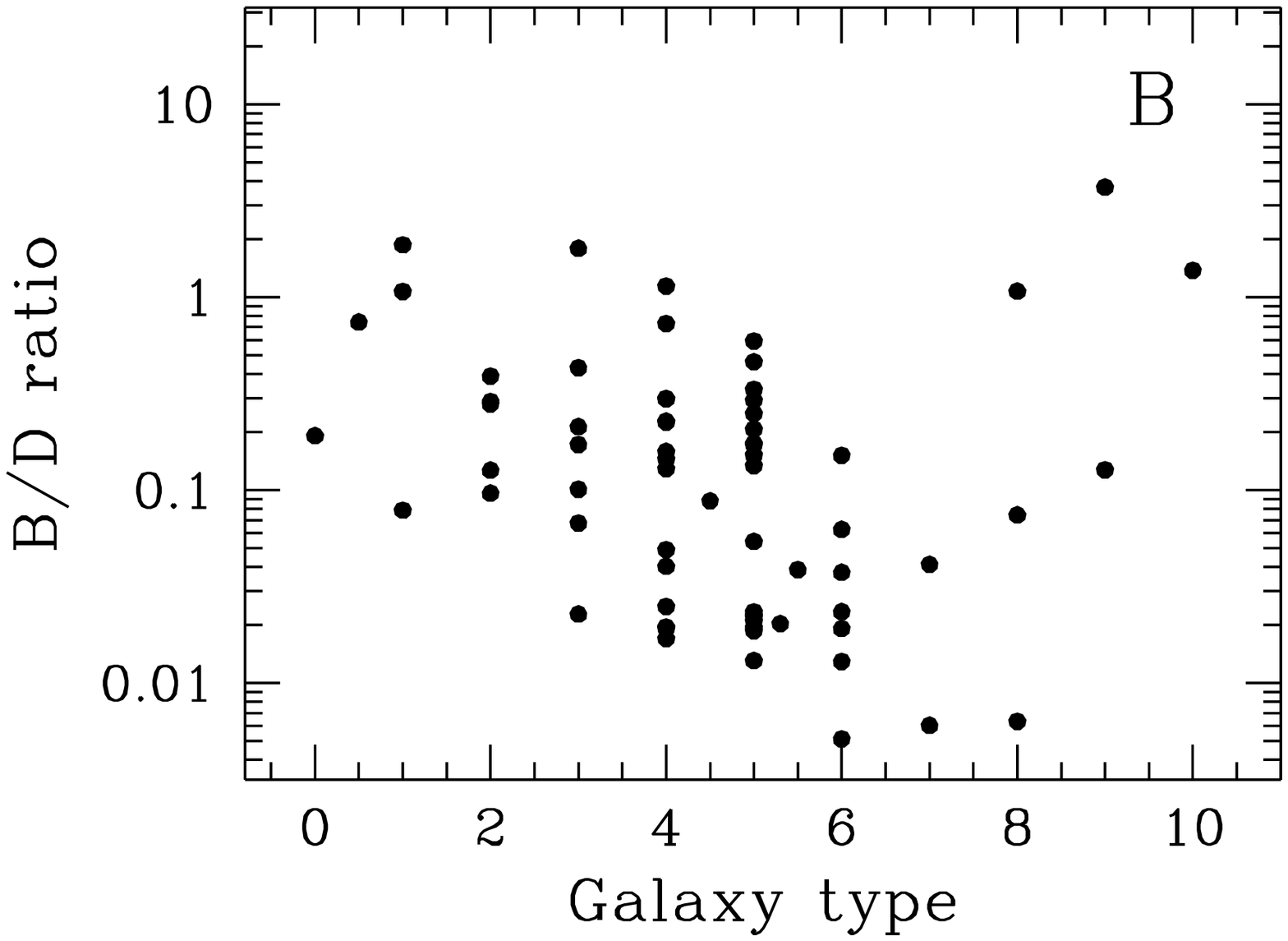}}
 \mbox{\epsfxsize=8.8cm\boundboxt{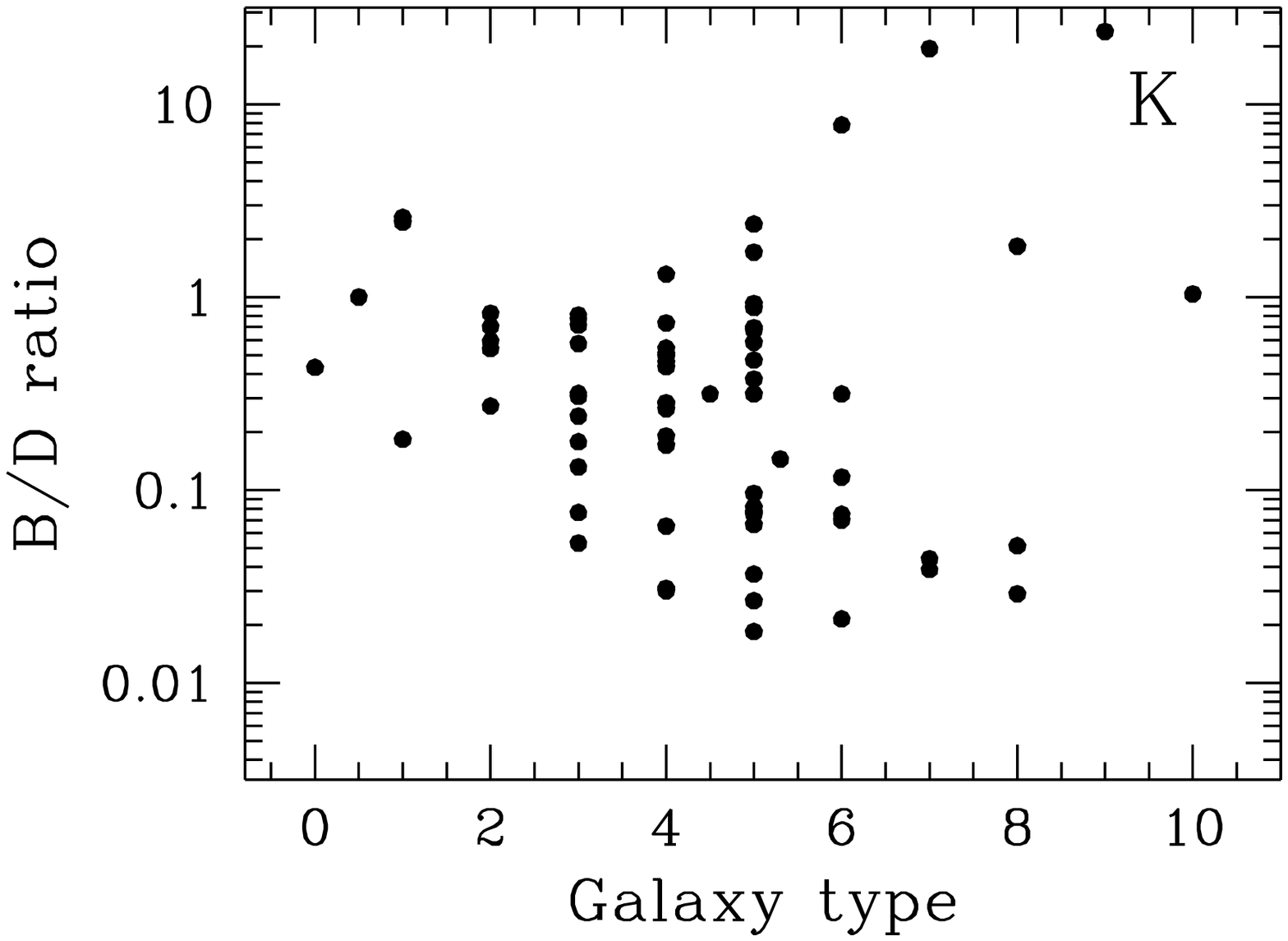}}
 \caption[]{
The bulge to disk ratio as function of morphological type using an \rq 
law bulge in the bulge/disk decomposition.
 }
 \label{type_lBD4}
 \end{figure*}

Comparing the $\muo^i$ with \mue\ (Fig.~\ref{cses}) we see no
correlation, except that the late-type spirals clearly stand out.  This
is most obvious in the $K$ passband.  The correlation between $h$
and \re\ (Fig.~\ref{scer}) is only tight in the $K$ passband and not in
the $B$ passband.  Actually the correlation is becoming steadily
tighter from the $B$ to the $K$ passband with correlation coefficients
increasing from 0.6 in $B$ and $V$ to $\sim0.75$ in $R$ and $I$ and
$\sim0.8$ in $H$ and $K$ passbands.  The equation for the least squares
fitted line is in the $K$ passband
 \begin{equation}
\log(r_{\rm e}^K) = 0.95 \log(h^K) - 0.86
\label{hrerel}
 \end{equation}
 with a standard deviation of 0.17.  The scalelength difference
between bulge and disk thus is of order 10.
 The relation also holds for all the 1D fit techniques presented in
Paper~II (but less strongly) except for the case of an \rq law bulge. 
In the case of 1D \rq law bulge there is at best a weak trend
(correlation coefficient 0.15). 
 This relation could partly be produced by the fitting routine if the
errors in both parameters are intrinsically correlated.  
The facts that this relation holds in the $K$ but not in
the $B$ passband and for both of the totally different 1D and 2D
fitting techniques indicate that the correlation is intrinsic and not
an artifact of the fit routines. The $\chi^2$ distribution around the
solutions found by the fit routine also showed no correlation with the
relationship between disk scalelength and bulge effective radius.

The absolute magnitudes of bulge and disk correlate well
(Fig.~\ref{MdMb}).  This is probably an example that large galaxies have
more of everything, more bulge, more disk.  The combined effect of good
correlation between \re\ and $h$ and weak correlation between \muo\ and
\mue\ ensures the correlation between $M_{\rm d}$ and $M_{\rm b}$. 
Looking at the bulge-to-disk ratio as function of morphological type
(Fig.~\ref{type_lBD}), one sees that there is a correlation, but this
correlation is less strong than for instance the one of \mue\ with type
(Fig.~\ref{type_es}).  The B/D ratio is on average higher in $K$ than in
$B$; bulges are redder than disks.  This partly explains why the
correlations which involve bulge parameters are tighter in $K$.  The
bulge/disk decomposition is more easily performed when the bulge is
relatively brighter.

Fig.~\ref{type_lBD} shows also that the selected galaxies are
indeed disk dominated systems.  The B/D ratios plotted here are much
smaller than the ratios normally found in the literature. This is
mainly due to the use of an exponential bulge. In Fig.~\ref{type_lBD4}
the results with an \rq law bulge (Paper~II) are shown. The B/D ratios
are higher, but the scatter has increased and there still is no tight
correlation with morphological type.

\section{Discussion}
\label{discus3}

In this section I will place the previously described relations in
the context of the three topics of main interest: 1) Freeman's law, 2)
bivariate distributions and 3) the relation between Hubble
classification and the structural parameters. 
 %I will conclude this section by
%confronting some galaxy formation and evolution theories with
%the newly found and some well known relationships. A combination of several
%models is probably needed to explain all aspects discussed here.

\subsection{Freeman's law}

Since Freeman~\cite{Freeman} found disk central surface brightnesses to
be constant among spiral galaxies, a number of explanations have been
brought forward.  In the introduction three explanations were
mentioned: 1) optically thick dust, 2) erroneous measuring of the disk
parameters from the light profiles and 3) selection effects. For each of
these possibilities I check if they are of importance for the current
sample and whether they can explain Freeman's result.

\subsubsection{Optically thick dust}
 It has been suggested that optically thick dust could be the cause of
Freeman's law (Jura~\cite{Jura}; Valentijn~\cite{Val90};
Peletier et al.~\cite{Pel94}).  This is only a partial explanation, because
it removes the inclination dependence from the law. To produce
Freeman's law in this way, galaxies must have the same surface
brightness at $\tau\! =\!1$ (where the typical surface brightness is
produced in an optically thick system), which means the problem is only
shifted from one part of the galaxy to another.  One now has to explain
why all galaxies have the same surface brightness at $\tau\!=\!1$. 
Taking a constant dust-to-stellar light ratio will not produce a
constant surface brightness.  This is only the case if all dust is in
front of the star light, but dust and stars are mixed in a galaxy and a
fraction of stars to the near side is less obscured.  The amount of
extinction in a galaxy is not linearly dependent on the amount of dust
present (de Jong~\cite{deJthes}). Coupling the amount of dust and stars in
galaxies can only reduce the scatter in \muo, but can never produce a
constant \muo.

I have shown that applying the inclination correction of
Eq.~(\ref{inccor}) reduces the scatter in the \muo\ of the disk going
from $C\!=\!0$ to $C\!=\!1$ (Table~\ref{avecsb}), indicating transparent
behavior.  The effect is small and the scatter is still dominated by
the intrinsic differences in the brightnesses of the disks.  One should
realize that the disk parameters are largely determined by
the outer regions of the galaxy.  They probably do not reflect the
optical thickness of the central regions. I note again that galaxies
can behave optically thin in an inclination test, while in fact being
optically thick. 

The $K$ passband data should hardly be affected by dust extinction.
Looking at Table~\ref{avecsb} one can see that the standard deviation
of \muo\ is for the early-type galaxies smaller in the $K$ passband
than in the $B$ passband, contrary to what is expected for optically
thick dust. The increase in standard deviations for the later types can
be explained by stellar population differences (Paper~IV). I conclude
that dust extinction is not a major effect in Freeman's law, certainly
not in the $K$ passband data used here.

\subsubsection{Erroneous profile fitting}
 Kormendy (\cite{Kor77}), Phillipps \& Disney (\cite{PhiDis83}) and
Davies (\cite{Dav90}) have argued that Freeman's law results from
fitting exponential disks to light profiles without taking the bulge
contribution to the profiles into account.  To prove their point, they
created model profiles with \rq law bulges and exponential disks with a
range of properties, to which exponential disks were fitted in a
specified range.  These models were able to reproduce Freeman's
``universal'' central surface brightness value of 21.65 $B$-\magarc\
with just a small scatter. 

The parameter space explored in the models is not very physical
according to current insights. Kormendy~(\cite{Kor77}) used B/D ratios of
1--120 for the low surface brightness systems and Phillipps \&
Disney~(\cite{PhiDis83}) assumed that bulges were so extended that they
dominated the light profiles again at the 24.5 $B$-\magarc.  Davies
(\cite{Dav90}) used a range of properties for the bulge parameters
which are typical in samples of galaxies, which suffer from severe
selection effects.  He used a {\em constant} central surface brightness
of the {\em bulge} to show that the central surface brightnesses of
disks need not be constant.  A change in $r_{\rm e}/h$ ratio was used
to produce a range in bulge-to-total light ratios ($BT$).  I have shown
that \mue\ is not constant (Fig.~\ref{type_es}) and that the $r_{\rm
e}/h$ ratio is nearly constant (Fig.~\ref{scer}).  Even though these
results where obtained with a exponential rather then a \rq law bulge,
a constant central brightness for bulges is excluded and a relationship
between \re\ and $h$ might be expected.

\begin{figure}
 \mbox{\epsfxsize=8.6cm\boundboxo{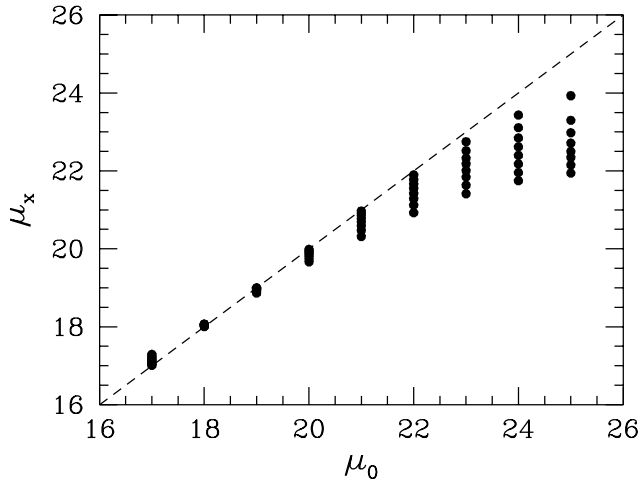}}
 \caption[]{
 The variation of the measured extrapolated central surface brightness
($\mu_x$) with bulge-to-total light ratio ($BT$) ranging from 0.05 to
0.75, at a given model central surface brightness \muo\ (see text). 
 }
 \label{davfig}
 \end{figure}

I use the parameterization of Davies~(\cite{Dav90}) to show that erroneous
bulge/disk decomposition is not a major factor in the Freeman law.
Figure~\ref{davfig} was produced in the same way as Davies' Fig.~6 by
fitting exponential profiles in the range of 22 to 25.5 \magarc\ to
model profiles, which were a combination of an \rq law bulge and an
exponential disk.  In this figure the intrinsic \muo\ of the model
profile is compared with the central surface brightness of the fitted
disk ($\mu_x$).  The $r_{\rm e}/h$ ratio was taken fixed to 0.4 and
$\mu_{\rm e}$ was adjusted to produce $BT$ ratios in the range from
0.05 to 0.75 in steps of 0.1 (contrary to Davies, who used a fixed
\mue\ and varied the $r_{\rm e}/h$ ratio).  The tendency to shift
intrinsic bright disks to the observed value of 21.65 \magarc\ has
disappeared.  The low surface brightness disks have too bright $\mu_x$
values for their \muo, but these were clearly fitted in the curved part
of the profile at the brighter end.  From Fig.~\ref{davfig} one can
conclude that it is unlikely that the high surface brightness disks
were underestimated (even with using the ``marking the disk'' fit of
Paper~II).  The situation for LSB systems is not as bad as it seems,
because the B/D ratios are low for LSB galaxies and the curvature of
the profiles can be readily seen.  Choosing values for \re/$h$ in the
range from 0.1 to 1 hardly changes these results. 

The central surface brightnesses of the galaxies were determined in
Paper~II with a full 2D decomposition technique, but also with the
``marking the disk'' technique and a few 1D decomposition techniques. 
If Freeman's result was caused by his use of the ``marking the disk''
method, the central surface brightnesses obtained with this method
should show large and systematic differences with the results of the
other methods.  In Paper~II it was shown that making a proper
decomposition of the profiles in a \rq law bulge and an exponential
disk adds a scatter of at most 0.4 \magarc\ to \muo\ with respect to
the ``marking the disk'' results.  Assuming that about the same value
would hold for Freeman's sample, this would still result in a rather
small range in \muo\ for his sample.  In Paper~II it is furthermore
argued that the 2D decomposition technique, using exponential light
profiles for both disk and bulge, is more accurate and reduces the rms
difference between the ``marking the disk'' method and the 2D fit to
0.3 \magarc. Both the model decompositions using Davies' method and the
comparisons between different decomposition methods of real galaxies
indicate that it is unlikely that Freeman's results were caused by
improper decompositions.

\subsubsection{Selection effects}
 After taking selection effects into account, Fig.~\ref{hiscs} shows
that there is no such thing as a simple Freeman's law for
galaxies with scalelengths larger than 1\,kpc.  There seems to be a clear
upper limit to the central surface brightness, which cannot be
explained by selection effects.  Even taking the apparent upper limit
in absolute luminosity in Fig.~\ref{sclcs} into account, there still
should have been galaxies with \muo\ brighter than 20 $B$-\magarc\ (16
$K$-\magarc) in the sample according to this figure. 
Figure~\ref{hiscs}  shows further that there is no strong decrease
in the number of galaxies with lower surface brightness.  The distribution
becomes narrower if we exclude late-type spirals, but this exclusion can
hardly be justified.  Late-type spirals are in many respects no
separate class of galaxies, but just a continuation of the trends set
by the earlier type spirals.  A clear example of such a trend is seen
in Fig.~\ref{type_es}.  

In Paper~II it was shown that there is at most $\sim$0.30 \magarc\ rms
uncertainty in the central surface brightnesses.  The uncertainties
also showed no correlation with the surface brightness itself
(Paper~II, Fig.~4) and the results presented here can therefore not be
the results of measurement errors.  The central surface brightness
distribution of Fig.~\ref{hiscs} changes in some details if one of the
other fit techniques of Paper~II is used, but the general trend remains
unchanged.  The same holds true when the distances of the galaxies are
calculated with other flow models.

Sample selections are influenced by both \muo\ and $h$ and therefore the
most important distribution for disk \mbox{dominated} galaxies is the bivariate
distribution in the (\muo,$h$)-plane (Fig.~\ref{bilogsccs}).  These two
parameters describe a large fraction of the light of disk dominated
galaxies and to derive this distribution one needs distances.  To derive
the distribution of \muo\ of sample of galaxies without knowing the
distance to the galaxies, one must assume that the distribution of \muo\
is not correlated to for instance $h$ or $M$ (Davies et
al.~\cite{Dav94}; McGaugh et al.~\cite{McG95}).  The total \muo\
distribution can only be calculated in this statistical way if the \muo\
distribution at each $h$ or at each $M$ has the same shape. 
Figure~\ref{bilogsccs} shows that this is probably not the case for $h$
and Figure~\ref{bilogMcs} shows the same for $M$.  The statistical
methods can at best only be used to get an impression of the \muo\
distribution.

\subsection{Bivariate distributions}

The reason for our limited knowledge of low surface brightness galaxies
is clearly indicated by the selection lines in Fig.~\ref{sclcs}.  The
use of catalogs like the UGC prevents galaxies with central surface
brightness fainter than $\sim$25 $B$-\magarc\ from being included in a
sample, independent whether the sample is \mbox{diameter} or magnitude
selected.  Only the use of deeper photographic plates enabling a
selection at fainter isophotes (Schombert et al.~\cite{Sch92}) or deep
CCD surveys will result in samples with a larger number of low surface
brightness galaxies.  A galaxy like Malin~I (Impey \&
Bothun~\cite{ImpBot89}), with $\muo\!\simeq\!26.5$ $B$-\magarc\ and
$h\!\simeq\!55$\,kpc, is not found in conventional catalogs, even though
it has an integrated magnitude comparable to M\,101 and a $\sim$10 times
as a large scalelength as M\,101.  Figure~\ref{bilogsccs} indicates that
galaxies like Malin~I are probably not very numerous in the local
universe, but this cannot be said of galaxies with
$\muo\!>\!24$~$B$-\magarc\ and $h\!\approx\!1$ kpc.  There is a clear
need for deeper local surveys, especially in the near-IR. 

\begin{figure*}
 \mbox{\epsfxsize=8.8cm\epsfbox[87 140 482 440]{pbfMcs.cps}}
 \mbox{\epsfxsize=8.8cm\epsfbox[67 140 462 440]{pkfMcs.cps}}
 \caption[]{
 The volume corrected bivariate distribution of galaxies in the
($\mu_0$,$M$)-plane.  The number density $\Phi(\muo^i$,$M$) is per bin
size, which is in steps of 1 mag in $M$ and 1 \magarc\ in $\muo^i$. 
 }
 \label{bilogMcs}
 \end{figure*}

\begin{figure*}
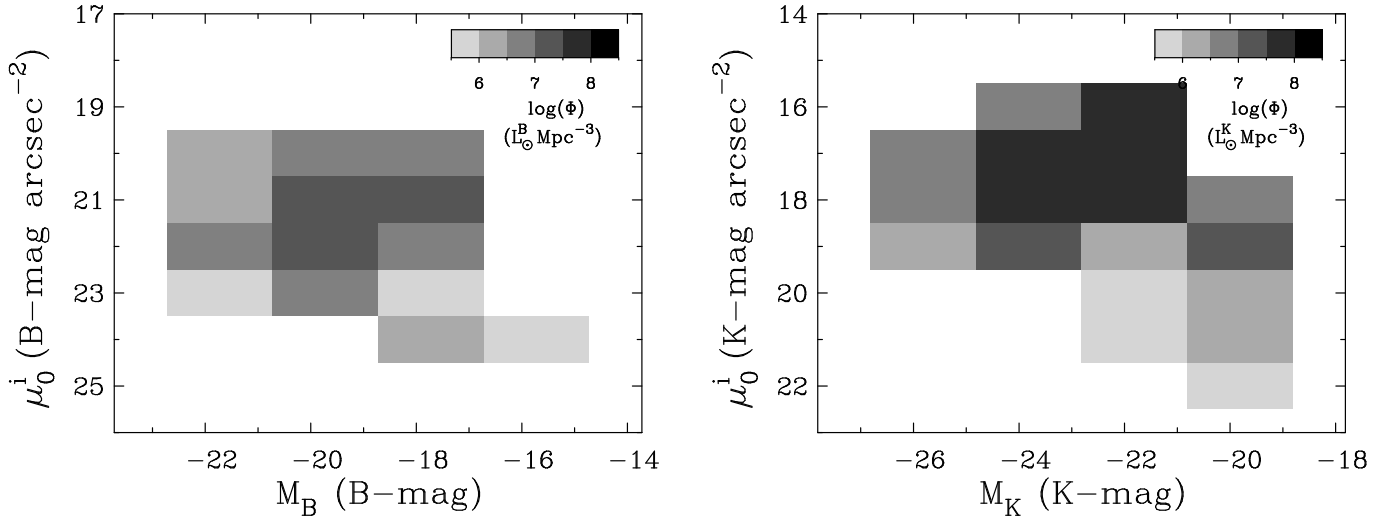

 \mbox{\epsfxsize=8.8cm\epsfbox[87 140 482 440]{pbfMcsw.cps}}
 \mbox{\epsfxsize=8.8cm\epsfbox[67 140 462 440]{pkfMcsw.cps}}
 \caption[]{
 The volume corrected luminosity distribution of galaxies in the
($\mu_0$,$M$)-plane. Each galaxy in Fig.~\ref{bilogMcs} was given an
additional weight according to its absolute luminosity.  The luminosity
density $\Phi(\muo^i$,$M$) is per bin size, which is in steps of 1 mag
in $M$ and 1 \magarc\ in $\muo^i$. 
 }
 \label{bilogMcsw}
 \end{figure*}

Few bivariate distributions have appeared in the literature which can
be used in comparisons with the current results. Van der Kruit
(\cite{Kru87}) calculated a bivariate distribution of spiral galaxies
in the (\muo-$h$)-plane. His distribution shows similar features as the
distribution presented here. The distribution has an upper limit in
central surface brightness at about 21 $J$-\magarc (photographic $J$
passband, which is similar to the Johnson $B$ passband) and an
exponentially declining density distribution with scalelength. 

Another comparison can be made with the bivariate distribution of van
der Kruit (\cite{Kru89}).  This distribution was constructed in exactly
the same way as Fig.~\ref{bilogsccs}, with only a modification to
Eq.~(\ref{vmaxcor}) to include the effects of an additional magnitude
selection limit.  The central surface brightnesses were convolved with
a Gaussian of 0.3 \magarc\ to incorporate the effects of calibration
uncertainty.  This explains the smoothness of the distribution.  The
galaxies of type later than T$\ \!=\ \!$5 were excluded and a Hubble
constant of 75 km s$^{-1}$ Mpc$^{-1}$ was used.  The same upper limits
in absolute magnitude and central surface brightness can be seen as in
Fig.~\ref{sclcs}.  There is only one galaxy with \muo\ brighter than
18~$R$-\magarc\ at 0.7(100/75) kpc.  The general trend in the bivariate
distribution of van der Kruit agrees quite well with
Fig.~\ref{bilogsccs}, with equal probabilities along lines of equal
absolute luminosity. 

The luminosity function of galaxies is a tool often used to investigate
galaxy evolution on cosmological time scales.  This makes sense as the
total luminosity of a galaxy seems, also by use of the TF-relation,
related to the total mass of a galaxy.  In determining the local LF one
has sometimes failed to notice that galaxies are extended and that for
selection correction one should not treat them as point sources.  This
can lead to a change in the slope of the LF as function of redshift and
results in the faint blue galaxy problem (McGaugh~\cite{McG94}). 
Another disadvantage of the one-dimensional LF is that each luminosity
bin contains galaxies with totally different surface brightnesses and
scalelengths (Fig.~\ref{sclcs}).  The physical processes in disk
galaxies seem to be more related to surface brightness than to total
mass (see also Paper~IV).  Therefore, to investigate the distributions
related to both the total mass and the surface brightness, the
luminosity function has been divided in several central surface
brightness bins in Fig.~\ref{bilogMcs}.  The absolute magnitudes were
calculated from the values given in Paper~I.  Figure~\ref{bilogMcs}
shows the bivariate distribution of a local sample of galaxies and can
be useful as a reference for high redshift samples observed with the
Hubble Space Telescope. 

Figure~\ref{bilogMcs} shows that the LF is more or less the same for
all central surface brightness bins in the $B$ passband.  In the $K$
passband something becomes apparent which was already hinted at in the
$B$ passband.  The LF for fainter central surface brightnesses is lower
and/or shifted to lower absolute luminosities.  Figure~\ref{bilogMcs}
is somewhere in between both options presented in Fig.~3 of
McGaugh~(\cite{McG94}), which means that both the shape of the LF and
its normalization depend on the bivariate distribution of $M$ and \muo.
Unfortunately, this data is too scarce to make a firm quantitative
statement, but it is clear that further attention should be given to
``the LF''. 

%It is also interesting to know what type of galaxies provides
%most of the total luminosity in the local universe. 
 At least as important as the number density of the local universe is
the local luminosity density, which can be transformed in a luminous
mass density using an $M/L$ description.  In order to calculate the
luminosity density, the number density value of each galaxy used in
Fig.~\ref{bilogMcs} was given an additional weight depending on its
absolute luminosity, which results in Fig.~\ref{bilogMcsw}.  This figure
shows the total luminosity one expects to find in a random Mpc$^3$ from
galaxies in the indicated bins of (\muo,$M$).  The luminosities are
expressed in solar luminosities per passband, calibrated using the
absolute solar luminosity values of Worthey (\cite{Wor94}). 

The distribution in the $B$ passband is remarkably flat, almost all
bins that contain galaxies are equal to within 1.5 order of magnitude.
In the $K$ passband distribution there is more structure, as most of the
$K$ passband light in the local universe comes from higher surface
brightness galaxies. Due to the scarceness of the data, this
figure is again more of qualitative than of quantitative interest.

\subsection{Hubble classification}

The Hubble sequence is one of the basic ingredients of galaxy formation
and evolution schemes, even though the underlying physical processes
are only partly understood. In this section I describe the consequences
of some of the relations between the structural parameters and Hubble
type as presented here.

Figure~\ref{type_lBD} showed that B/D ratio cannot be used to determine
the Hubble type of face-on systems, which means that the classification
of edge-on systems is different from that of face-on systems. 
Furthermore, the B/D ratios seem to be quite small ($<0.5$), even in
the $K$ passband where the differences in color (and $M/L$) due to
stellar population effects between disk and bulge are minimized.  The
B/D ratios are larger when \rq law bulges are used.  The difference
between exponential and \rq law bulges in terms of generalized
exponential profiles is extensively discussed in Paper~II.  Young \&
Curry (\cite{YouCur94}) showed that for ellipticals and dwarf
ellipticals there is a trend in profile shape with luminosity.  The
brighter ellipticals have \rq like profiles, while fainter (dwarf)
ellipticals have more exponential like profiles.  If this is also true
for bulges, we might expect early-type spirals to have more centrally
peaked bulge profiles than later type spirals.  In Paper~II it was
found that \rh law bulges gave smaller $\chi^2$ residuals for T$<$3
than exponential bulges.  This picture conflicts with the scale
independence of Hubble type as seen in the $K$ passband data of
Fig.~\ref{scer}.  Within one Hubble type, a range in integrated bulge
luminosities exists, which should result in different profiles when the
model of Young \& Curry (\cite{YouCur94}) is applied to bulges.  There
is a weak indication for this trend, because the galaxies with the
brightest bulges are slightly better fitted by the \rh law bulges than
by the exponential bulges. The \rq law bulges never give the smallest
residuals, not even for the most luminous bulges.

Using an exponential bulge, an important relation is found, which is
not apparent if an \rq law bulge is used: $h$ correlates with \re, but
these parameters do not correlate with Hubble type (Fig.~\ref{scer}). 
%The correlation between $h$ and \re\ will not hold if other bulge
%profile functions are used.  Still t
 This is an important relation as it makes the Hubble sequence scale
free.  Each Hubble type comes in a range of different sizes, both in
terms of diameter and total luminosity.
 %(and therefore by the TF-relation in a range of maximum rotation velocities, $V_{\rm rot}$). 
 It is an example of larger (in the sense of scale size, not mass)
galaxies having more of everything, larger disk, larger bulge.  Due to
this correlation the scale parameters cancel each other out in
calculating B/D ratios, which means that a plot of \mue--\muo\
($\propto \Sigma_{\rm e}/\Sigma_0$, important in density wave models)
versus type looks like Fig.~\ref{type_lBD} with a different scaling. 
Consequently \mue--\muo\ is also a bad diagnostic for Hubble type. 

The relationship between Hubble type and \mue\ (Fig.~\ref{type_es})
holds, independent of bulge profile function used.  In fact the
relation holds if one just fits a line to the central region of the
profile and uses the true central surface brightnesses, because the
bulge generally dominates the luminosity in this central region. 
One might wonder if in classifying galaxies one has mainly looked at
the surface brightness of the bulge and not at the B/D ratio.  For the
earlier systems this is harder to accept; the central regions are in
general overexposed on photographic plates used for classifying. It is
instructive to know that S0 galaxies do not fit in this relation.  The
central surface brightnesses of S0 galaxies range from $\sim$17.7 to
$\sim$19 $B$-\magarc, estimated from the data of Kormendy
(\cite{Kor77}) and Peletier et al.~(\cite{Pel90}). This translates to
effective surface brightnesses in the range of 19.5-21~$B$-\magarc, in
accordance with the two S0 galaxies in the current sample, but
significantly below the trend of the rest of the spiral galaxies
(Fig.~\ref{type_es}). 
%The difference is smaller at the longer
%wavelengths (earlier types have redder centers, Paper~IV), but never
%vanishes.

If B/D ratio is such a bad diagnostic for Hubble type, we are left
according to Sandage~(\cite{San61}) with only two other classification
discriminators: 1) the spiral arm structure, 2) the pitch angle of the
arms.  The two remaining criteria indicate that the Hubble sequence
should be explained in terms of spiral arm appearance, even though the
second criterion might also be in doubt, as measurements by Kennicutt
(\cite{Ken81}) showed that pitch angle has no tight correlation with
morphological type.
 % and that $V_{\rm rot}$ might be a second parameter here. 

The main theory on spiral structure is the spiral density wave theory
(Lin \& Shu~\cite{LinShu64}; Roberts et al.~\cite{Rob75}).  The fact
that Hubble type is a scale free quantity fits into this theory.  When
the bulge and the disk scale with the same amount, the shape of the
rotation curve stays the same, only its amplitude changes.  This means
that the shape of the resonances also scale along with the scalelength
changes. If bulge and disk scalelengths are correlated, the shape of
the rotation curve is fully determined by the relative brightness of
the bulge and disk component. Therefore it is harder to understand in
the density wave theory, why Hubble type does not correlate tightly with
B/D ratio (Fig.~\ref{type_lBD}), or to be more precise with \mue--\muo.  The
strength and the pitch angle of the density wave gets modified by the
ratio of mass that participates in the density wave to the mass that
does not.  If B/D ratios are so small that they hardly could affect the
density wave (Fig.~\ref{type_lBD}) and if on top of that the B/D ratios
and \mue--\muo\ values are only weakly correlated with Hubble type, it
seems that some modifications to the standard density wave model are
needed.  Maybe a connection between the \mue\ of the bulge and the
distribution of dark matter can solve this problem.

\section{Conclusions}
\label{concl3}
 The statistics of the fundamental parameters of 86 spiral galaxies have
been studied in the optical and the near-IR.  The use of the near-IR $K$
passband enabled for the first time determination of these parameters
without being hampered by the effects of dust extinction and differences
in stellar populations.  Volume density distributions with respect to
the fundamental parameters were made, which was possible due to the
careful selection of the sample.  The main conclusions are as follows:

\begin{itemize}
 \item Freeman's law of a preferred disk central surface brightness
value needs a modification.  Although there seems to be a clear upper
limit to the central surface brightnesses of galaxies, there is no
clear limit at the faint end of the \muo\ distribution.  The number of
galaxies in a volume with a certain \muo\ is only slowly declining
function of \muo.  

\item The bulge and disk scalelengths are correlated, parameterised by
$\log(r_{\rm e}^K) \!= \!0.95 \log(h^K) \!- \!0.86$. This correlation
suggests that the formation of the bulge and the disk is coupled.

\item The Hubble classification is related to the surface brightness
properties of spiral galaxies.  However, the relations are in general
not very tight and can therefore not be turned around to give
morphological classification.  The B/D ratio is not a good indication of
Hubble type (Fig.~\ref{type_lBD}~and~\ref{type_lBD4}).  The best
relation with Hubble type found in this study is the one with \mue\
(Fig.~\ref{type_es}).  The physical interpretation is difficult, because
cause and effect are hard to separate. 

\item Hubble type is a scale size independent parameter, but not a total
luminosity independent parameter of a galaxy.  Therefore, it would be
better to divide by scale size instead of luminosity to derive scale
independent parameters of galaxies in comparisons.  The suggestion that
Hubble type is mainly driven by total mass (Zaritsky~\cite{Zar93}) seems
an oversimplification.  To carry this point a bit further, it is
probably better to separate the determination of the LF of galaxies into
bins which are related to the effective surface brightness (like
Fig.~\ref{bilogMcs}) than into bins which are related to Hubble type. 
 \end{itemize}

Larger samples are needed to enable the parameterization of the
bivariate distributions. There is especially need for accurate surface
photometry of a large sample of galaxies, selected from deep
photographic plates providing isophotal diameters at very faint levels.

 \begin{acknowledgements}
 I thank Ren\'ee Kraan-Korteweg for providing her Virgo-centric inflow
computer model. Piet van der Kruit and Reynier Peletier are thanked for
the many fruitful discussions. Erwin de Blok, Ren\'e Oudmaijer, David
Sprayberry and Arpad Szomoru are acknowledged for their many useful
suggestions on the manuscript.  
This research was supported under grant
no.~782-373-044 from the Netherlands Foundation for Research in
Astronomy (ASTRON), which receives its funds from the Netherlands
Foundation for Scientific Research (NWO).
 \end{acknowledgements}

\end{document}